\documentclass{article}
\usepackage{arxiv}

\usepackage[utf8]{inputenc} 
\usepackage[T1]{fontenc}    
\usepackage{hyperref}       
\usepackage{url}            
\usepackage{booktabs}       
\usepackage{amsfonts}       
\usepackage{nicefrac}       
\usepackage{microtype}      
\usepackage{lipsum}		
\usepackage{graphicx}
\usepackage{natbib}
\usepackage{doi}
\usepackage{amssymb}
\usepackage{amsmath}
\usepackage{amsthm}
\usepackage{mathrsfs}
\usepackage{algorithm}
\usepackage{algpseudocode}
\usepackage{subcaption}
\usepackage{graphicx}

\newtheorem{theorem}{Theorem}
\newtheorem{remark}{Remark}
\newtheorem{lemma}{Lemma}

\title{Decentralised Q-learning for Multi-Agent Markov Decision Processes With a Satisfiability Criterion}

\date{} 					

\author{Keshav P. Keval\\
	Department of Electrical Engineering, IIT Bombay\\
	 Mumbai-400076\\
	\texttt{keshavpatel564@gmail.com} \\
	\And
	Vivek S. Borkar \\
	Department of Electrical Engineering, IIT Bombay\\
	Mumbai-400076\\
	\texttt{borkar.vs@gmail.com} \\
}
	
\renewcommand{\shorttitle}{Decentralised Q-learning for MMDPs With a  Satisfiability Criterion}

\hypersetup{
pdftitle={Decentralised Q-Learning for Multi-Agent MMDPs with a Satisfiability Criterion}, 
pdfsubject={eess.SY},
pdfauthor={Keshav P. Keval, Vivek S. Borkar},
pdfkeywords={Decentralised learning, gossip algorithm, Metropolis-Hastings scheme, multiplicative weights update, multi-agent control, Q-learning, replicator equations, satisfiability},
}

\begin{document}
\maketitle

\begin{abstract}
In this paper, we propose a reinforcement learning algorithm to solve a multi-agent Markov decision process (MMDP). The goal, inspired by Blackwell's Approachability Theorem, is to lower the time average cost of each agent to below a pre-specified agent-specific bound. For the MMDP, we assume the state dynamics to be controlled by the joint actions of agents, but the per-stage costs to only depend on the individual agent's actions. We combine the Q-learning algorithm for a weighted combination of the costs of each agent, obtained by a gossip algorithm with the Metropolis-Hastings or Multiplicative Weights formalisms to modulate the averaging matrix of the gossip. We use multiple timescales in our algorithm and prove that under mild conditions, it approximately achieves the desired bounds  for each of the agents. We also demonstrate the empirical performance of this algorithm in the more general setting of MMDPs having jointly controlled per-stage costs.
\end{abstract}


\section{Introduction}
\label{sec: intro} 
This paper operates within the decentralised paradigm for multi-agent learning, where the agents communicate over a communication network, represented by an undirected graph. Each agent communicates with its neighbours on the graph, sharing local information such as actions, per-stage costs and observations (for the partially observable case). In this section we motivate the need for such a decentralised setting by stating some problems inherent to multi-agent learning that can be remedied by this setting.
\subsection{Scalability}
\label{sub: Scalability}
Unlike with single-agent learning, when multiple agents are present in the environment, from the perspective of each agent, the environment appears to be non-stationary. 
In dealing with the non-stationarity of the environment, each agent needs to either keep track of/model the actions taken by the other agents in order to make updates to their Q-values. This makes naive Q-learning infeasible for a large number of agents, since the dimensionality of the Q-function varies exponentially with the number of agents. This can be remedied if each agent only needs to keep a track of ``local" information while learning. This motivates the decentralised paradigm where communication takes place through a gossip-like protocol, allowing the information from one part of the graph to eventually ``percolate" to the rest.
\subsection{Information Structures}
\label{sub: Information structures}
In the presence of multiple agents, the question "Who knows what?" becomes important. It is, in general, not reasonable to assume that
each agent knows the actions/policies of the other agents, since the agents interests can be dis-aligned. This also aggravates the non-stationarity faced by each agent, since samples of actions from opponents cannot always faithfully recover the policy being followed by these agents. One possible remedy for this is to assume the presence of a centralised controller, which collects information such as joint action, joint costs or joint observations (in case of partial observability) and can design policies for agents. In many cases the presence of such a centralised controller becomes unrealistic, due to a lack of robustness or perhaps due to
the communication overhead of communicating with a central unit. This paves the way for the decentralised paradigm, which is more realistic and also makes theoretical analysis easier than in the general setting, where no information is exchanged between agents.

\section{Overview of Classical Q-learning for Multi-Agent Learning problems}
The multi-agent settings can be broadly classified into two types based on whether each agents knows the ``local" information of the other agents. When we assume that each agent knows the actions of the rest, the setting is called Joint Action Learning. When the agents do not know that actions of the rest of the agents, the setting is called Opponent Modelling. We highlight the use of Q-learning inspired algorithms to perform joint action learning in stochastic games. A stochastic game $(N,S,A,R)$ consists of $N$ agents, a set of states $S$, the set of joint actions $A = \Pi_{i \in [N]} A_{i}$, where $A_{i}$ is the set of actions of player $i$ and a reward function $R:S\times A \to \mathbb{R}$. The following algorithm jointly describes the Q-learning schemes that we will discuss.

\begin{algorithm}
\caption{Q-learning for Stochastic games}\label{alg:2}
\textbf{Initial Conditions}\\
$Q^{i}(s,a) = 0, \forall (s,a) \in S\times A, i \in [N].$\\
Repeat for all episodes:\\
For $t = 0,1,2, \cdots$ :
    \begin{itemize}
        \item Observe the current state $s_{t}$.
        \item With probability $\epsilon$ choose a random action from this state for all agents.
        \item Otherwise solve $\Gamma_{s_{t}}$ to get the joint equilibrium policy $[\pi^{*,1}_{t},\pi_{t}^{*,2},\cdots,  \pi_{t}^{*,N}]$ and take action $a_{t}^{i} \sim \pi_{t}^{*,i}$.
        \item Observe the joint action $[a_{t}^{1}, a_{t}^{2}, \cdots a_{t}^{N}]$, the rewards of each agent $r_{t}^{1}, r_{t}^{2}, \cdots , r_{t}^{N}$ and the next state $s_{t+1}$.
        \item For all agents $j=1$ to $N$:\\
        $$Q_{t+1}^{j}(s_{t}, a_{t}^{j}) = Q_{t}^{j}(s_{t}, a_{t}^{j}) + \alpha[r_{t}^{j} + \gamma \cdot \text{Value}_{j}(\Gamma_{s_{t+1}}) - Q_{t}^{j}(s_{t},a_{t}^{j})].$$
    \end{itemize}
\end{algorithm}
In the above algorithm, $\Gamma_{s_{t}}$ is a single-stage static game where the reward of player $i$ for the joint action $a$ is $Q_{t}^{i}(s_{t}, a)$. To compute the equilibrium $\pi^{*}_{t}$ of this game, we need an appropriate game-theoretic notion of optimality such as Nash equilibrium, correlated equilibrium or security. Once we settle on a notion of optimality, at every iteration we calculate the "Value" of $\Gamma_{s_{t}}$ as : 
$$ \text{Value$_{i}$}(\Gamma_{s_{t}}) = \sum_{a \in A}Q^{i}_{t}(s,a) \pi^{*}_{t}(a|s),$$
where $\pi^{*}_{t}(a|s)$ is the probability of joint action $a$ being taken by the agents in the state $s$ under the equilibrium policy $\pi^{*}_{t}$.
We then use it in the Q-learning update equation. In case of single-agent MDP, "Value" would just be a max operation over the actions of an agent.
For a detailed survey of multi-agent learning see \citet{MARL-survey} and \citet{marl-book}.
We now present a few algorithms with different notions of optimality for finding the equilibrium policy:
\begin{itemize}
    \item \textbf{Minimax Q-learning} \citep{minimax-Q-learning}: This algorithm can be used in two-player zero-sum stochastic games, it can calculate the minimax value of the game.  In this case, the "Value" is computed by finding the value obtained by the minimax policy for the game $\Gamma_{s_{t}}$ at each iteration $t$. This can be considered as the Q-learning version of the value iteration algorithm proposed for two-player zero-sum stochastic games in \citet{shapley}.
    \item \textbf{Nash Q-learning} \citep{nash-Q-learning}: 
    This algorithm can be applied to general-sum stochastic games, and we obtain convergence under some strict assumptions. We compute the equilibrium policy for $\Gamma_{s_{t}}$ by finding the Nash equilibrium.
    \item \textbf{Correlated Q-learning} \citep{correlated-Q-learning}: 
    This algorithm can also be applied to general-sum stochastic games. We solve the game $\Gamma_{s_{t}}$ by finding the correlated equilibrium. Note that it may not be possible to write a correlated equilibrium policy in terms of individual agent policies. In this case, the agent samples a joint action from the equilibrium policy and just plays the individual action prescribed by that joint action.
    \item \textbf{Friend-or-Foe Q-learning} \citep{friend-foe-Q-learning}:  
    In this algorithm, we assume that at the beginning of the learning process, every agent designates each of the other agents as either a "Friend" or "Foe". Each agent then calculate the "Value" of the game as follows:
    \begin{equation*}
    \begin{split}
    \text{Value$_{i}$($\Gamma_{s}$)} = \max_{\pi \in \Pi(X_{1}, \cdots, X_{k})}\min_{y_{1}, \cdots, y_{l}\in Y_{1}\times\cdots \times Y_{l}}\sum_{x_{1},\cdots,x_{k} \in X_{1}\times\cdots \times X_{k}}\pi(x_{1})\cdots\pi({x_{k}}) \cdot  Q_{i}(s, x_{1}, \cdots, x_{k}, y_{1}, \cdots, y_{l}), 
    \end{split}
    \end{equation*}
    where $X_{1}$ through $X_{k}$ are the friends and $Y_{1}$ through $Y_{l}$ are the foes of agent $i$ respectively. Intuitively, agent $i$ assumes that its friends are cooperating with it and the foes are conspiring against it to lower its reward. The Friend-or-Foe Q-learning algorithm converges even in a general sum stochastic game, though the values learned may not correspond to any Nash equilibrium of the game.
\end{itemize}
\section{Overview of Decentralised Multi-Agent Algorithms}
The decentralised paradigm focuses on settings where the agents communicate with each other over a communication graph (network). Each agent makes decisions based on local information, which includes information conveyed by the neighbours on a graph. Such a setting has applications in robotics, sensor networks, intelligent transportation systems and smart grids. In addition to being robust, compared to the centralised setting, the decentralised schemes save on the computational costs that would have been incurred at some central processing unit. This is especially the case as the number of agents become large.\\
We now present a  few algorithms in this setting, to give a big picture idea of the tasks that can be accomplished in this paradigm. For a survey of learning in this paradigm, see \citet{DMARL-survey}.


\subsection{QD-Learning}
An example with cooperative decentralised agents is as follows: consider the task of optimizing the infinite horizon discounted networked average reward:
$$ Arg\max_{\pi \in \Pi}\mathbb{E}_{\pi}\left [ \frac{1}{N}\sum_{i=1}^{N}\sum_{t\geq1}\gamma^{t-1}R(X_{t}, \pi(X_{t}))\right],$$
where $N$ is the number of agents, $R(\cdot,\cdot) : S \times A^{N}$ maps state-joint action pairs to rewards and $\Pi$ is the set of joint policies. 
The following multiple timescales update is run by each agent $i$:
\begin{equation*}
\begin{split}
Q^{i}_{t+1}(s,a) = Q^{i}_{t}(s,a) + \alpha_{t,s,a}\left[ R^{i}(s,a) + \gamma\max_{a' \in A^{N}}Q^{i}_{t}(s',a') - Q^{i}_{t}(s,a)\right]  - \beta_{t, s, a}\sum_{j \in \mathscr{N}_{t}^{i}}\left[Q^{i}_{t} - Q^{j}_{t}(s,a)\right],
\end{split}
\end{equation*}
where $\alpha_{t,s,a}, \beta_{t,s,a}$ are step-sizes chosen appropriately and $\mathscr{N}_{t}^{i}$ is the set of neighbours of agent $i$ on the communication network at time $t$. The above update equation can be broken into consensus+innovation parts, with the terms multiplying $\beta_{t,s,a}$ responsible for asymptotic consensus and the terms multiplying $\alpha_{t,s,a}$ responsible for innovation, i.e. incorporating new reward information into the optimal Q-value estimate. This scheme provably converges to the joint policy which maximizes the infinite horizon networked average reward, for the tabular setting and in the case of a finite state-action space. For proof of convergence see \citet{QD-learning}.


\subsection{Decentralised Actor-Critic Algorithm}
With the increase in the size of the joint-action space, function approximation becomes necessary to tackle scalability. This leads naturally to policy-based approaches such as the decentralised version of the Actor-Critic algorithm. If agent $i$'s policy is parameterised by $\theta_{i}$, $\theta = [\theta_{1}, \theta_{2}, \cdots \theta_{N}]$, let $Q_{\theta}$ be the global Q-function corresponding to the networked average infinite horizon discounted reward. One can prove that the policy gradient with respect to agent $i$'s parameter is:
$$\nabla_{\theta_{i}}J(\theta) = \mathbb{E}[\nabla_{\theta_{i}}\log \pi^{i}_{\theta^{i}}(s,a^{i})Q_{\theta}(s,a)],$$
where $\pi^{i}_{\theta^{i}}(s,a^{i})$ is the probability that agent $i$ takes the action $a^{i}$ from state $s$ under the parameterized policy $\pi^{i}_{\theta^{i}}$.
Since each agent cannot estimate the global Q-function from just the local information, each agent keeps a copy $Q_{\theta}(\cdot, \cdot, \omega^{i})$ of the global Q-function and runs the following consensus based TD-learning update:
$$ \tilde{\omega}_{t}^{i} = \omega_{t}^{i} + \beta_{\omega, t}\delta_{t}^{i}\nabla_{\omega} Q_{t}(\omega_{t}^{i}),$$
$$ \omega_{t+1}^{i} = \sum_{j \in \mathscr{N}}c_{t}(i,j)\tilde{\omega}_{t}^{j},$$
where $\beta_{\omega,t}$ is an appropriate step-size, $\delta_{t}^{i}$ is the local TD-error calculated using $Q_{\theta}(\cdot, \cdot, \omega^{i})$ and $c_{t}(i,j)$ are edge weights over the communication network which is doubly stochastic, to ensure asymptotic consensus. For the actor step, each agent performs stochastic policy gradient using its own copy of the global Q-function. For details see \citet{zhang-decentralized-actor-critic}


\subsection{Decentralised Fitted Q-iteration Algorithm}
This is a batch-based algorithm, where all the agents collaborate to iteratively update the estimates of the global Q-function (again, for the infinite horizon networked average reward). This is done by fitting non-linear least squares with $y_{j}^{i} = r_{j}^{i} + \gamma \max_{a' \in A}Q_{t}^{i}(s_{j}',a'), j=1 \text{ to }n$ as the responses ($Q_{t}^{i}$ is agent $i$'s Q-function estimate at time $t$). Assuming that $\mathscr{F}$ is the class of non-linear functions that we are fitting over, the joint goal for the agent then becomes to solve the following distributed optimization problem :
$$ \min_{f \in \mathscr{F}}\frac{1}{N}\sum_{i =1}^{N}\frac{1}{2n}\sum_{j=1}^{n}[y_{j}^{i} - f(s_{j}, a_{j}^{1}, a_{j}^{2}, \cdots a_{j}^{N})]^{2}.$$
Further, searching over a class of functions $\mathscr{F}$ which makes the objective convex for each $i$ allows us to use familiar distributed optimization techniques to solve for the global minima.
For more details, see \citet{Fitted-Q-iteration}

\subsection{Mean Square Projected Bellman Error}
As opposed to a learning algorithm, this is a policy evaluation algorithm which can be used in the multi-agent setting. The policy evaluation task in the multi-agent scenario is much cleaner than the learning task, since in the absence of learning by all agents, the environment faced by each of the agents becomes stationary. For the joint policy $\pi$, let the agents parameterize their value function by $\{ V_{\omega}(s):=\phi^{T}(s) \omega : \omega \in \mathbb{R}^{d}\}$, where $\phi(s)$ is the feature vector at state $s$ and $\omega$ is the parameter vector. Let us define $D$ as the $|S| \times |S|$ diagonal matrix with the state occupancy measures as the diagonal elements and define $\Phi = [\cdots; \phi^{T}(s); \cdots] \in \mathbb{R}^{|S|\times d}$. Then the goal becomes to jointly solve the following optimization problem:
$$ \min_{\omega}||\Pi_{\Phi}(V_{\omega} - \gamma P^{\pi}V_{\omega} - R^{\pi})||^{2}_{D},$$
where $\Pi_{\Phi}$ is the projection onto the column space of $\Phi$, $P^{\pi}$ is the transition probabilities under the joint policy $\pi$ and $R^{\pi}$ is the infinite horizon networked average reward. Using Fenchel's Duality Theorem, one can convert the above problem into a distributed saddle point problem. This problem can then be attacked by either the original formulation or the saddle-point formulation. For a variety of work that uses the above method, see \citet{DMARL-survey}.

\subsection{Beyond Rewards}
When stepping into the multi-agent world, many other notions of optimality and performance criteria arise. This includes the communication efficiency of the agents. One can reduce the number of communication rounds between the agents and a central controller by designing parsimonious communication trigger rules.

\section{Problem Statement}
There are $N$ agents communicating over a network, represented by a undirected connected graph $G = (V,E)$. Let $\mathscr{N}(i) := \{j \in V: (i,j) \in E\}$ represent the set of neighbours of agent $i$ in the graph. Let $\{X_{n}\}$ be a controlled Markov chain, controlled by the joint action of the $N$ agents. Without loss of generality, let each agent $i \in [N]$ have the same action space $A$. The transition function is :
$p(\cdot, \cdot,\cdot) : S\times A^{N}\times S \to [0,1].$
The objective is to find a joint stationary policy $\pi^{*} = [\pi^{*, 1}, \cdots \pi^{*,N}] : S \to A^{N}$, among the set of joint stationary policies $\Pi$ such that the following inequality is satisfied for each agent $i \in [N]$:
\begin{equation}\label{eq:satisfiability}
    \limsup_{T \to \infty}\frac{1}{T}\sum_{t = 1}^{T}\mathbb{E}_{\pi^{*}}\left[c^{i}(X_{t}, \pi^{*,i}(X_{t})\right] \leq \beta^{i},
\end{equation}
where $\mathbb{E_{\pi}}$ denotes the expectation under the joint stationary policy $\pi$. The constants $\beta^{i}$ and cost functions $c^{i}(\cdot, \cdot): S\times A\to \mathbb{R}$ are prescribed for each agent $i \in [N]$. Note that the per-stage cost of agent $i$ depends only on the its own action.
The goal is to come up with an algorithm that yields a joint stationary policy which satisfies \eqref{eq:satisfiability}.
We make the following assumptions :\\
\textbf{Assumption 1} : The Markov chain is irreducible under any joint stationary policy.\\
\textbf{Assumption 2} : There exists a policy in $\Pi$ which satisfies \eqref{eq:satisfiability}.\\

The problem statement is inspired by the concept of Blackwell approachability \citep{Blackwell}, which is a generalization of the Minimax theorem for two player zero sum games where the payoffs are vectors rather than scalars. The goal of one of the players in such an "approachability" game is to drive the time average of vector payoffs to a convex set, while the opponent tries to prevent this from happening. Blackwell provided a necessary and sufficient condition for this in repeated games. 
This is also related to the work on Q-learning for Multi-criterion Markov Decision Processes by \citet{Q-satisfiability}.

\section{Preliminaries}
In this section, we introduce the meta-algorithms used for our proposed solution to the problem.

\subsection{Multiple Timescale Stochastic Approximation}
Consider the following coupled iterations:
$$ x_{n+1} = x_{n} + a(n)[h(x_{n}, y_{n}) + M^{1}_{n+1}], $$
$$ y_{n+1} = y_{n} + b(n)[g(x_{n}, y_{n}) + M^{2}_{n+1}]$$
We assume that the step-size sequence $a(n)$ and $b(n)$ satisfy the Robbins-Monro conditions: 
$$ a(n), b(n) > 0 , \sum_{n}a(n) = \infty, \sum_{n}b(n) = \infty, \sum_{n}(a(n))^{2} + (b(n))^{2} < \infty.$$
In addition to this, we assume that the function $h$ and $g$ are both Lipschitz, in order to maintain the well-posedness of the corresponding ordinary oifferential equations (ODEs) for the above iterations. $M^{i}_{n}$ for $i=1,2$ is a square-integrable Martingale noise sequence. For the above iterations, we also assume that the step size sequence satisfy:
 $$ \frac{b(n)}{a(n)} \to 0.$$
 This means that the step-size $b(n)$ decays to zero much faster than $a(n)$, which puts the $x_{n}$ iteration on a faster timescale and the $y_{n}$ iteration on a slower timescale.
 Now consider the singularly perturbed ODE: 
 $$ \Dot{x}(t) = \frac{1}{\epsilon}h(x(t), y(t)),$$
 $$ y(t) = g(x(t), y(t)).$$
 In the limit $\epsilon \to 0$, the first ODE acts as a fast transient and the second ODE acts as a slower component, appearing ``almost stationary" to the faster component. The $x_{n}$ and $y_{n}$ iteration behave in a similar manner, whereby the $x_{n}$ iteration appears ``quasi-stationary" to the $y_{n}$ iteration and the $y_{n}$ iteration appears to be ``quasi-static" to the $x_{n}$ iteration. \\
To highlight the usefulness of multiple timescales, consider an iterative algorithm which calls another iterative algorithm during every iteration. For every iteration of the outer loop, one has to wait for the inner iteration/loop to terminate or converge. This has to be done for every run of the outer loop and makes the algorithm somewhat episodic.\\
If we instead run the two iterations concurrently with different step-sizes such that one decays to zero faster than the other, then we get the same effect as described above. Thus, we can introduce multiple timescales into an algorithm without making it episodic. This will become important later, since we use this argument to justify the convergence of Q-learning in a Multi-agent environment.\\
 We now formalize the notion of two timescales. 
 Consider the $y_{n}$ iteration, but re-written in the following form :
 $$ y_{n+1} = y_{n} + a(n)[\epsilon_{n} + M^{3}_{n+1}],$$
$$ \epsilon_{n}:= \frac{b(n)g(x_{n}, y_{n})}{a(n)},\,\, M^{3}_{n+1} = \frac{b(n)M^{2}_{n+1}}{a(n)}.$$
Note that $\epsilon_{n}$ is a  bounded $o(1)$ sequence and can be ignored while analysing the ODE for the above iteration.
We can now treat the above iteration as running on the same timescale as $x_{n}$, but with $g(\cdot, \cdot)$ replaced by $0$. Hence the $y_{n}$ iterations ``track" the ODE: 
$$ \Dot{y}(t) = 0.$$
By "track", we mean that the iterates $y_{n}$ converges to a (possibly sample path dependent) internally chain transitive invariant set of the above ODE. 
We can similarly conclude that the $x_{n}$ iterations "track" the ODE: 
$$ \Dot{x}(t) = h(x(t), y(t)).$$
This gives us : $(x_{n}, y_{n})$ jointly "track" the ODE: 
$$ \Dot{x}(t) = h(x(t), y(t)), $$
$$ \Dot{y}(t) = 0.$$
Stronger results on the convergence of two timescale stochastic approximation exist, but are not stated here since the above is enough to understand the motivation behind the use of multiple timescales. See \citet{borkar2009stochastic-book}, Chapter 6 for a detailed coverage of this topic.

\subsection{Gossip Algorithm}
Let there be $N$ agents that are communicating over a connected undirected graph $G = (V,E)$, where the vertices represents the agents and the edges are communication links, assumed to be reliable. Each agent $i$ maintains an iterate $\{y^{i}_{n}\}$ and based on the iterates of its neighbours $\mathscr{N}(i)$ on the graph $G$, makes the following update:
$$ y^{i}_{n+1} = \sum_{j \in \mathscr{N}(i)\cup\{i\}}\Tilde{p}(j|i)y^{j}_{n} + a(n)(h^{i}(y_{n}) + M_{n+1}^{i}),$$
where $[[\Tilde{p}(j|i)]]$ is an irreducible stochastic matrix, which we call the Gossip matrix, $\{a(n)\}$ is a positive step-size sequence that satisfies the Robbins-Monro conditions and $M_{n}^{i}$ for all $i$ is a Martingale noise sequence. The iterates asymptotically ``track" the following ODE:
$$\dot{y}^{i}(t) = \sum_{i = 1}^{N}\Tilde{\pi}(i)h^{i}(y(t)), \,\, \forall i \in [N], $$
where $\Tilde{\pi}$ is the stationary distribution associated with the Gossip matrix $\Tilde{p}$. The iterates of all agents converge to a common limit, which is the stationary point of the above ODE. See \citet{Gossip} and \citet{non-linear-gossip} for a proof and generalization respectively.

\subsection{Stability Criterion for Stochastic Approximation}

Consider the following iterations : 
$$ x_{n+1} = x_{n} + a(n)[h(x_{n}) + M_{n+1}].$$
We do not assume the boundedness of the iterates to begin with. When the iterates cross/escape the unit ball around the origin we re-scale the iterates to put them back inside the unit ball with a scaling factor $c\geq 1$. We can perform this re-scaling in a periodic manner after every $T$ steps. This periodic re-scaling gives rise to scaled iterates $\hat{x}_{n}$ and the scaled function :
$$ h_{c}(x) := \frac{h(cx)}{c}, c \geq 1.$$
Consider the case when the iterates are unbounded, this means that the iterates (scaled) keep leaving the unit ball, which in turn implies that the scaling factor $c$ required to keep the iterates within the unit ball goes to $\infty$. Then the limiting  ODE is :
$$ \dot{x}(t) = h_{\infty}(x(t)),$$
$$ h_{\infty}(x) = \lim_{c \to \infty}h_{c}(x).$$
If this limiting ODE has the origin as a unique globally asymptotically stable equilibrium, then the iterates will have to fall at an exponential rate into the unit ball. If the iterates escape to infinity, the size of the jumps from within the unit ball to outside have to become unbounded. By using Gronwall's inequality, one can prove that this is not possible. \\
Hence, if the origin is a unique globally asymptotically stable equilibrium of the limiting ODE $\dot{x}(t) = h_{\infty}(x(t))$, then the iterates remain bounded.\\
This is just one possible stability test, which happens to be useful for us. See \citet{borkar2009stochastic-book}, Chapter 3 for more on stability criteria. 

\subsection{The Multiplicative Weights Update Rule}

The Multiplicative Weights Update (MWU) rule is a generalization of the Weighted Majority algorithm from learning theory. It is a general meta-algorithm which can be used to solve a variety of problems. It has many applications: solving two-player zero-sum games approximately, finding $\mathcal{O}(\log(n))$ approximations to many NP-Hard problems as well as boosting in Machine Learning.
We describe the MWU in a general framework of finding, in an online manner,  the action $l^{*}$ among a set of $L$ actions with the lowest cost.\\
Let $[L]$ be the set of actions,  and $c(l)$ be the cost incurred when action $l$ is selected. In order to recognize the best action, we randomize actions using  a probability distribution over the set of actions $p_{t}(l)$ for $l \in [L]$ and update it based on the cost incurred at each step as follows: 
$$ p_{t+1}(l) = \frac{p_{t}(l)(1 - \gamma)^{c(l)}}{\sum_{j}p_{t}(j)(1 - \gamma)^{c(j)}},$$
where $\gamma$ is small positive constant. This rule is sometimes called the "Hedge" rule. For details about MWU and further applications see \citet{Arora2012TheMW}.
\subsection{The Replicator Equations}
The Replicator Equations evolve within the $d$-dimensional probability simplex: 
$$ \Delta_{d} := \left\{w = [w_{1}, w_{2}, \cdots w_{d}]^{T} : \sum_{i = 1}^{d}w_{i} = 1, w_{i} \geq 0 \,\, \forall  i \in [d]\right\},$$
and are given by:
$$ \dot{w}_{i}(t) = w_{i}(t)\left[D^{i}(w(t)) - \sum_{j=1}^{d}w_{j}(t)D^{j}(w(t))\right],$$
where $w_{i}$ is the fraction of the $i'$th species in the population and $D^{i}(\cdot) : \mathbb{R}^{d} \to \mathbb{R}$ is the ``fitness"/``payoff" of the $i'$th species. In the convergent case, the dynamics eventually concentrates on the species with the maximum ``fitness"/``payoff" i.e. the other species become extinct. For a detailed exposition see \citet{hofbauer_sigmund_1998}, Chapter 7.
\\
The "Hedge" rule described previously is the discrete time version of  the following replicator equation:
$${\dot{p}_{t}(l)} = p_{t}(l)\left((1-\gamma)^{c(l)} - \sum_{k}p_{t}(k)(1-\gamma)^{c(k)}\right).$$
This replicator equation and the multiplicative weights update have the same stationary points. In this case, both the schemes would concentrate on the action with the lowest cost.
We use a modification of this rule in our proposed algorithm.

\subsection{The Metropolis-Hastings Scheme}
We can use the Metropolis-Hastings scheme \citep{Metropolis1953EquationOS, Hastings} to sample from: 
$$ \pi(i) \propto \exp\left({-\frac{V(i)}{T}}\right), \,\, T>0, i \in [m].$$
To do this, we sample from a random walk on a graph $G = (V,E)$, with $V = [m]$ and some $E$ that makes the graph connected. We use the following transition matrix : 
$$ p(j|i) = \left\{
\begin{array}{ll}
       \frac{1}{\text{deg}(i)}\exp\left(-\frac{(V(j) - V(i))}{T}^{+}\right),& j\neq i, j \in \mathscr{N}(i)\\
      0, & j \not \in \mathscr{N}(i)\cup\{i\}\\
      1 - \sum_{j\in \mathscr{N}(i)}p(j|i), & j=i. \\
    \end{array}\right.
$$
For a modern treatment of MH see \citet{robert2016metropolishastings}.

\subsection{Relative Q-Factor Iteration}
Consider an an average reward Markov decision process (MDP) $M= (S,A,P,C)$, where $S$ is the state space, $A$ is the action space, $P:\nolinebreak S\times A \times S \to [0,1]$ is the transition function and $C:S\times A \to \mathbb{R}$ is the per-stage cost function.
Fix constants $s_{0} \in S, a_{0} \in A$ of the MDP. The asynchronous relative Q-factor iterations for an average cost MDP with per stage cost $c(\cdot,\cdot)$ are given by:
\begin{equation*}
\begin{split}
Q_{n+1}(s,a) = Q_{n}(s,a) + \gamma(v(n,s,a))[C(s,a) + \min_{b}Q_{n}(\xi^{s,a}_{n},b)  - Q_{n}(s_{0},a_{0}) - Q_{n}(s,a)] \mathbb{I}\left((s,a) \in Y_{n}\right),  
\end{split}
\end{equation*}
where, 
\begin{itemize}
    \item $\{\xi^{s,a}_{n}\}_{n \geq 1}$ are i.i.d. random variables such that $\mathbb{P}(\xi^{s,a}_{n} = s') = p(s'|s,a), \,\,\forall (s,a,s')\in S\times A \times S.$
    \item The positive step size $\{\gamma(n)\}$ satisfies the Robbins-Monro conditions:
    $\sum_{n}\gamma(n) = \infty, \sum_{n}\gamma(n)^{2} < \nolinebreak\infty$
    \item $\{Y_{n}\}$ is a set-valued process taking values in the subsets of $S\times A$, defined by:
    $$ Y_{n} := \{(s,a) : Q_{n}(s,a) \text{ is updated at time $n$}\}.$$
    \item $v(n,s,a) :=$ the number of times the pair $(s,a)$ has been updated till time $n$, i.e., 
    $$v(n,s,a) := \sum_{m=1}^{n}\mathbb{I}\left((s,a) \in Y_{m}\right), \,\, \forall n\geq 1.$$
\end{itemize}
Due to the asynchronous nature of the updates, we need the following assumptions :\\
\textbf{Assumption 3} Let $[(\cdot)]$ denote ``the integer part of $(\cdot)$", then for any $x\in(0,1)$ the step size $\gamma(\cdot)$ satisfies :
$$ \sup_{k}\frac{\gamma([xk])}{\gamma(k)} < \infty \, \, \text {and}\,\,\, \frac{\sum_{m=0}^{[yk]}\gamma(m)}{\sum_{m=0}^{k}\gamma(m)} \to 1\,\, \text{uniformly for $y \in [x,1]$}$$\\
\textbf{Assumption 4}
There exists $\delta>0$ such that 
$$ \liminf_{n \to \infty}\frac{v(n,s,a)}{n+1} \geq \delta \,\, \text{a.s.}$$
with $(s,a) \in S\times A$. Furthermore, for all $x>0$ and 
$$ N(n,x) = \min\{m\geq n : \sum_{k = n}^{m}\gamma(k)\geq x\},$$
the limit 
$$ \lim_{n \to \infty}\frac{\sum_{k = v(n,i,a)}^{v(N(n,x), i, a)}\gamma(k)}{\sum_{k = v(n,j,a)}^{v(N(n,x), j, a)}\gamma(k)}\,\,\, \text{exists a.s. for all $i,j,a,u$}.$$
For a justification of the above assumptions, see \citet{abounadi}.
\\
The update equation for the relative Q-factor iterations "track" the following ODE :
$$ \Dot{Q}_{t}(s,a) = G(Q_{t}) - Q_{t},$$
where the operator $G : \mathbb{R}^{|S| \times |A|} \to \mathbb{R}^{|S| \times |A|}$, is defined as follows :
$$ G(Q)(s,a) = C(s,a) + \sum_{s'}p(s'|s,a)\min_{b}Q(s\,b) - Q(s_{0}, a_{0}).$$

\section{Intuition behind the Proposed Algorithm}

Since we only want to "drive" the average joint behaviour of the agents such that all the agents have an average cost less than their pre-specified thresholds, our setting is close to the "approachability" landscape, in that we merely want to drive the joint policy to within the set of policies where the required asymptotic bounds are satisfied (This set is guaranteed to be non-empty due to Assumption 1).

We heavily use the multiple-timescale idea that was briefly discussed in the preliminaries, in this case the proposed algorithm has three algorithmic timescales as well as an update running on the natural timescale. On the slowest timescale, we run a Gossip iteration over the communication graph, in order to allow the agents to share information with each other. The Gossip matrix for this algorithm is itself modulated in the natural timescale. This modulation of the Gossip matrix can take place either through the Multiplicative Weights Update or the Metropolis-Hastings scheme. The modulation of the Gossip matrix is done to put the maximum weight on the agent which is violating its bound by the most. Due to the averaging property of the Gossip iteration, eventually all the agents achieve consensus on which agent is performing the worst, and they each adjust their actions in order to lower the costs of this worst performing agent. Each agent also runs an iteration on the middle algorithmic timescale to keep a track of their time average cost. On the fastest algorithmic timescale we run a relative Q-factor iteration with the per-stage costs equal to the gossip iterate. Once the gossip iteration achieves consensus about the worst performing agent, the Q-learning of each agent makes them adapt their actions to make the environment more "favourable" for the worst performing agent.

\section{Algorithm}
A brief description of each of the timescales is given below:
\begin{itemize}
    \item \textbf{Natural Time Scale}: This is update rule \eqref{eq:MWU} and \eqref{eq:MH} and is inspired by the Multiplicative Weights Update rule and the Metropolis-Hastings scheme. It adjusts the weights that are used for averaging during the Gossip iteration \eqref{eq:gossip}. For each agent, the neighbouring agents whose running cost is greater than the specified threshold will get a higher weight. As a result, the stationary distribution of the Gossip matrix will "single out" the agent(s) that deviate the most from their desired bounds.
    \item \textbf{Fast Algorithmic Timescale}:
    This is \eqref{eq:Q-learning} and is meant to find the policy which each agent $i$ must follow in order to minimize the Gossip iterate $y^{i}_{n}(\cdot,\cdot)$, obtained from \eqref{eq:gossip} running on a slower timescale. We take $\gamma_{1}(n) = \frac{1}{n^{0.8}}$ in \eqref{eq:Q-learning}.
    \item \textbf{Medium Algorithmic Time Scale}:
    This is \eqref{eq:running-cost} and is used to keep track of the running average cost of each agent. We take $\gamma_{2}(n) = \frac{1}{n^{0.9}}$ in \eqref{eq:running-cost}.
    \item \textbf{Slow Algorithmic Time Scale}:
    This is \eqref{eq:gossip} and implements a Gossip algorithm, which tracks the weighted average of the costs of each of the agents, weighted by the stationary distribution of the Gossip matrix. The Gossip matrix is modulated on the natural timescale by either the Multiplicative Weights Update rule in \eqref{eq:MWU} or the Metropolis-Hastings scheme \eqref{eq:MH}. We take $\gamma_{3}(n) = \frac{1}{n}$ in \eqref{eq:running-cost}. 
\end{itemize}

\begin{remark}
In general, we want $\gamma_{1}(n),\gamma_{2}(n),\gamma_{3}(n)$ that each satisfy the Robbins-Monro conditions and $ \gamma_{2} = o(\gamma_{1}), \gamma_{3} = o(\gamma_{2})$. Additionally, due to asynchronous updates, $\gamma_{1}(n)$ and $\gamma_{3}(n)$ should satisfy Assumptions 3 and 4. 
\end{remark}
The algorithm is described on the next page.

\begin{algorithm}
\caption{Decentralised Q-learning for multi-agent MDPs with a satisfiability criterion}\label{alg:1}
Let $s_{0},a_{0}$ be arbitrary elements of $S$ and $A$ respectively. Select $T, \gamma, \epsilon, \epsilon_{w} >0.$ \\
\textbf{Initial Conditions}\\
$Q^{i}_{0}(s,a) = 0, \forall (s,a) \in S\times A, i \in [N]$\\
$ p_{0}(j|i) = \frac{1}{1 + \text{deg}(i)}, \forall j\in \mathscr{N}(i)\cup\{i\}, i \in [N] $\\
$\pi^{i}_{0}(s) \in Arg\min_{b}Q^{i}_{0}(s,b), \forall i \in [N], \forall s \in S$\\
$\pi_{0}(s) = [\pi_{0}^{1}(s), \pi_{0}^{2}(s), \cdots \pi_{0}^{N}(s)]^{T}$\\
$X_{0} = s_{0} \in S$\\
Use either \eqref{eq:MWU} or \eqref{eq:MH} throughout\\
\textbf{For} $n=1,2,3,\cdots $ \textbf{do} : \\
\begin{itemize}
    \item Generate set-valued random variable $Y_{n}^{i}$ for each agent $i$ and random variable $\xi_{n}^{s,a}$.
    \item For all $(s,a) \in S \times A$ and $i\in[N]$ set: 
    \begin{equation}\label{eq:Q-learning}
    \begin{split}
        Q_{n+1}^{i}(s,a) = Q_{n}^{i}(s,a) + \gamma_{1}(v(i,n,s,a))[y_{n}^{i}(s,a) + \min_{b \in A}Q^{i}_{n}(\xi_{n}^{s,a}, b) \\- Q_{n}^{i}(s_{0}, a_{0}) -  Q_{n}^{i}(s,a)]\mathbb{I}((s,a) \in Y_{n}^{i}))
    \end{split}
    \end{equation}
    \item For each agent $i\in [N]$ set :
    \begin{equation}\label{eq:running-cost}
    z_{n+1}^{i} = z_{n}^{i} + \gamma_{2}(n)(c^{i}(s,\pi_{n}^{i}(s)) - z_{n}^{i})
    \end{equation}
    \begin{equation}\label{eq:gossip}
    \begin{split}
    y_{n+1}^{i} (s,a)=
        \mathbb{I}((s,a) \in Y_{n}^{i})\left[\sum_{j\in \mathscr{N}(i)\cup\{i\}}p_{n}(j|i)y_{n}^{j}(s,a) + \gamma_{3}(v(i,n,s,a))(c^{i}(s,a) - y_{n}^{i}(s,a))\right]  \\ + y_{n}^{i}(s,a) (1 - \mathbb{I}((s,a) \in Y_{n}^{i}))
    \end{split}
    \end{equation}
    \begin{equation}\label{eq:MWU}
    \begin{split}
     p'_{n+1}(j|i) = \left\{
    \begin{array}{ll}
       p_{n}(j|i)(1+\gamma)^\frac{z_{n}^{j} - \beta^{j}}{T}, & z_{n}^{j} \geq \beta^{j}, j \neq i\\
      p_{n}(j|i)(1-\gamma)^\frac{\beta^{j} - z_{n}^{j}}{T}, & z_{n}^{j} < \beta^{j}, j \neq i\\
      p_{n}(i|i), & j=i \\
    \end{array} 
    \right. \\
    p_{n+1}(j|i) = (1 - \epsilon_{w})\frac{p'_{n+1}(j|i)}{\sum_{k}p'_{n+1}(k|i)} + \frac{\epsilon_{w}}{1 + |\mathscr{N}(i)|}, \; \forall j \in \mathscr{N}(i)\cup\{i\}
    \end{split}
    \end{equation}
    $$ \text{OR}$$
    \begin{equation}\label{eq:MH}
        p_{n+1}(j|i) = \left\{
\begin{array}{ll}
       \frac{1}{\text{deg}(i)}\exp\left(\frac{-((z_{n}^{i} - \beta^{i}) - (z_{n}^{j} - \beta^{j}))^{+}}{T}\right), & j\neq i, j \in \mathscr{N}(i)\\
      0 & j \not \in \mathscr{N}(i)\cup\{i\}\\
      1 - \sum_{j\in \mathscr{N}(i)}p(j|i), & j=i. \\
    \end{array}\right.
    \end{equation}
    \item Update the policy :
    $$ \forall i \in  [N], \,\, \pi_{n+1}^{i} \in Arg\min_{b}Q^{i}_{n+1}(s,b) \,\, w.p. \,\, (1-\epsilon) \text{ and Uniform(A)}\,\, w.p. \,\, \epsilon $$ 
    $$ \pi_{n+1} = [\pi_{n+1}^{1}, \pi_{n+1}^{2}, \cdots \pi_{n+1}^{N}]^{T}$$
    \item $n \leftarrow n+1$
\end{itemize}
\textbf{end For}
\end{algorithm}

\section{Convergence of the Algorithm}

In this section we assume that $\sup_{n}||Q^{i}_{n}|| < \infty $ for all agents $i$, we prove this in the next section. The main result of this section is to prove convergence of the policy to one that satisfies the required bounds approximately.
\subsection{Definitions}
Let $g$ denote the vector of average costs for each of the agents:
$$g : \pi \to g(\pi) = \lim_{T \to \infty}\frac{1}{T}\sum_{m=1}^{T}\mathbb{E}_{\pi}[c(X_{m}, \pi(X_{m}))],$$
where $\pi$ is a joint stationary randomized policy, $c(\cdot, \cdot)$ is the vector of per-stage costs and the expectation is taken component-wise.
Let $\Delta_{N}$ be the $N$-dimensional probability simplex and $\Delta_{N}^{\epsilon}$ be the truncated simplex:
$$ \Delta_{N}^{\epsilon} := \{w \in \mathbb{R}^{N}: \sum_{i}w(i) = 1, \epsilon \leq w(i) < 1 \,\, \forall i \in [N] \}.$$
\subsection{Technical results on Metropolis-Hastings and Multiplicative Weights}
We need the following results due to practical reasons. When implementing the Gossip algorithm, we need an irreducible gossip matrix. Practically, if any of the weights $p(j|i)$ become too small it will be treated as a zero and henceforth will remain zero. This can cause problems with the convergence of the Gossip iteration in some cases. The following results guarantee that the weights do not become too small. This comes with a trade-off, as we shall see later, when we do this we only get approximate optimality.
\begin{lemma}
The Metropolis-Hastings (MH) scheme \eqref{eq:MH} remains in the truncated probability simplex $\Delta_{N}^{\epsilon}$, for some $\epsilon(T)>0$, where $T>0$ is the temperature parameter. Additionally, as $ T \to 0$, $\epsilon(T) \to 0$ and the stationary distribution concentrates on the agent $i$ with the highest $V(i) = z^{i} - \beta^{i}$ .
\end{lemma}
\begin{proof}
The stationary distribution for \eqref{eq:MH} is given by:
$$ \pi(i) = \frac{\exp(V(i)/T)}{\sum_{j = 1}^{N}\exp(V(j)/T)}.$$
This distribution is positive and always lower bounded by some $\epsilon(T)>0$.
With decreasing temperatures, this stationary distribution concentrates on the component with the highest $V(i)$.
\end{proof}
\begin{lemma}
The Multiplicative weights update (MWU) \eqref{eq:MWU} has the continuous time equivalent:
\begin{equation}\label{eq:replicator_equiv}
\Dot{p}_{t}(j|i) = p_{t}(j|i)\left(m^{j}_{i} - \sum_{k}m^{k}_{i}p_{t}(k|i)\right) + \left(\frac{1}{1 + |\mathscr{N}(i)|} - p_{t}(j|i)\right),
\end{equation}
where the payoff $m^{j}_{i}$ for agent $j$ in agent $i$'s replicator equations is given by: 
$$ m^{j}_{i} = \left\{
\begin{array}{ll}
       \frac{z^{j} - \beta^{j}}{T},& j\neq i, j \in \mathscr{N}(i)\\
      0, & \text{otherwise} \\
    \end{array}\right.
$$

\end{lemma}
\begin{proof}
The proof follows along the same lines as the one given in \citet{Krichene:EECS-2016-156} and we prove this for agent $i$'s update scheme in the limit as $\gamma \to 0$.
Let $X^{j}_{i}(t)$ be $L^{1}$ functions for $j \in \mathscr{N}(i)\cup\{i\}$, such that the discretizations of these functions at times $\{T_{n}\}_{n\geq 1}$ correspond to the discrete time update given in \eqref{eq:MWU} for agent $i$: $X_{i}^{j}(T_{n}) = p_{n}(j|i)$. Let the discretization be such that $\epsilon_{n} := T_{n+1} - T_{n}$ is a decreasing non-summable sequence. Define $I_{1} = \{j \in \mathscr{N}(i): z^{j} \geq \beta^{j}\}$ and $I_{2} = \{j \in \mathscr{N}(i): z^{j} < \beta^{j}\}$. Define $M_{n}$ and let $L_{n}$ be its first order Taylor approximation in $\gamma$:  $$M_{n} = \sum_{k \in I_{1}}p_{n}(k|i)(1+\gamma)^{\frac{z^{k} - \beta^{k}}{T}} + \sum_{l \in I_{2}}p_{n}(l|i)(1 - \gamma)^{\frac{\beta^{l}-z^{l}}{T}} + p_{n}(i|i),$$
$$ L_{n} = 1 + \sum_{k \in I_{1}}\gamma p_{n}(k|i)\left(\frac{z^{k} - \beta^{k}}{T}\right) + \sum_{l \in I_{2}}\gamma p_{n}(l|i)\left(\frac{z^{l} - \beta^{l}}{T}\right) + o(\gamma).$$
Consider the expression for $ p_{n+1}(j|i) - p_{n}(j|i) = X_{i}^{j}(T_{n+1}) - X^{j}_{i}(T_{n})$ for $j\in I_{1}$: 

$$ X^{j}_{i}(T_{n+1}) - X^{j}_{i}(T_{n}) = (1 - \epsilon_{w})\left( \frac{X_{i}^{j}(T_{n})(1+\gamma)^{\frac{z^{j} - \beta^{j}}{T}}}{M_{n}} - X_{i}^{j}(T_{n})\right) + \epsilon_{w}\left( \frac{1}{1 + |\mathscr{N}(i)|} - X_{i}^{j}(T_{n})\right).$$
Taking the first order Taylor approximation, we get:
$$ X_{i}^{j}(T_{n+1}) - X^{j}_{i}(T_{n}) = (1 - \epsilon_{w})\left( \frac{X_{i}^{j}(T_{n})\left(1+\gamma\left(\frac{z^{j}-\beta^{j}}{T}\right)+ o(\gamma)\right)}{L_{n}} - X_{i}^{j}(T_{n})\right) + \epsilon_{w}\left( \frac{1}{1 + |\mathscr{N}(i)|} - X_{i}^{j}(T_{n})\right).$$
In the small $\gamma$ limit, we can write the following :
$$ L_{n}^{-1} = 1 - \gamma\left(\sum_{k \in I_{1}}X_{i}^{k}(T_{n})\left( \frac{z^{k} - \beta^{k}}{T}\right) + \sum_{l \in I_{2}}X_{i}^{l}(T_{n})\left( \frac{z^{l} - \beta^{l}}{T}\right)\right) + o(\gamma).$$
Using use this in the previous equation and ignoring $o(1)$ terms, we get:
$$\frac{X^{j}_{i}(T_{n+1}) - X^{j}_{i}(T_{n})}{T_{n+1} - T_{n}}\cdot \frac{\epsilon_{n}}{\gamma} = (1 - \epsilon_{w}) X_{i}^{j}(T_{n})\left( \frac{z^{j} - \beta^{j}}{T} - \sum_{k \in \mathscr{N}(i)}X^{k}_{i}(T_{n})\left(\frac{z^{k} - \beta^{k}}{T}\right) \right) + \frac{\epsilon_{w}}{\gamma}\left(\frac{1}{1 + |\mathscr{N}(i)|} - X_{i}^{j}(T_{n})\right).$$
Let $\epsilon_{n} = \epsilon_{w} = \gamma \to 0$, we get the  desired ODE limit:
$$ \Dot{X}^{j}_{i}(t) = X^{j}_{i}(t)\left( \frac{z^{j} - \beta^{j}}{T} - \sum_{k \in \mathscr{N}(i)}\left(\frac{z^{k} - \beta^{k}}{T}\right)X^{k}_{i}(t) \right) + \left(\frac{1}{1 + |\mathscr{N}(i)|} - X_{i}^{j}(t)\right).$$ 
The proof is similar for the cases when $j\in I_{2}$ and $j=i$. See \citet{hofbauer_sigmund_1998}, Chapter 7 for more details on the classical replicator equations.
\end{proof}
\begin{lemma}
The continuous time replicator equation \eqref{eq:replicator_equiv} has the simplex as an invariant set and does not have any stationary points on the faces of the simplex.
\end{lemma}
\begin{proof}
To see this, check that $\frac{d}{dt}\sum_{j}p_{t}(j|i) = 0, \,\forall i \in [N].$
Also, the derivative $\Dot{p}_{t}(j|i)$ is positive for $p_{t}(j|i) = 0$ which means there can be no stationary points on the faces of the simplex. 
\end{proof}
\subsection{Approximate Optimality}
We will prove that the saddle point of a certain function, defined over the set of policies, satisfies the required bounds for each agent approximately. The following lemma will be useful later.
\begin{lemma}{\label{lemma 1}}
    The following is true:
    $$ \inf_{\pi \in \Pi}\sup_{w \in \Delta_{N}}\sum_{i = 1}^{N}w(i)(g^{i}(\pi) - \beta^{i}) \leq 0.$$
\end{lemma}
\begin{proof}
    $$ \inf_{\pi \in \Pi}\sup_{w \in \Delta_{N}}\sum_{i = 1}^{N}w(i)(g^{i}(\pi) - \beta^{i}) = \inf_{\pi \in \Pi}(g^{i^{*}}(\pi) - \beta^{i^{*}}) \leq 0,$$
    where $i^{*}\in Arg\max_{i \in [N]}(g^{i}(\pi) - \beta^{i})$ and the inequality in the last step follows from the assumption that there exists a joint policy in $\Pi$ which satisfies the bounds for all the agents (Assumption 1).
\end{proof}

\begin{theorem}\label{thm : saddle-point}
Let $\epsilon < \frac{1}{N}$ and $(\pi^{*,\epsilon}, w^{*, \epsilon})$ be the saddle point of the function: $\Pi \times \Delta_{N}^{\epsilon} \to \sum_{i}w(i)(g^{i}(\pi) - \beta^{i})$ :
$$ \pi^{*,\epsilon} = Arg\min_{\pi \in \Pi} \max_{w \in \Delta_{N}^{\epsilon}}\sum_{i} w(i)(g^{i}(\pi) - \beta^{i}),$$ $$w^{*, \epsilon} = Arg\max_{w \in \Delta_{N}^{\epsilon}}\sum_{i}w(i)(g^{i}(\pi) - \beta^{i}).$$
Then $\pi^{*,\epsilon}$ satisfies the bounds $\{\beta^{i}\}$ approximately :  
$ \max_{i}\left(g^{i}(\pi^{*,\epsilon}) - \beta^{i}\right) \leq \nolinebreak \mathcal{O}(\epsilon).$
\end{theorem}
\begin{proof}
$$ \min_{\pi \in \Pi}\max_{w \in \Delta_{N}^{\epsilon}}\sum_{i=1}^{N}w(i)(g^{i}(\pi) - \beta^{i}) \leq \min_{\pi \in \Pi}\sup_{w \in \Delta_{N}}\sum_{i=1}^{N}w(i)(g^{i}(\pi) - \beta^{i})\leq 0.$$
We can explicitly write the l.h.s as:
$$ \sum_{j\neq i^{*}}\epsilon\left(g^{i}(\pi^{*,\epsilon}) - \beta^{i}\right) + \left(1 - (N-1)\epsilon\right) \left(g^{i^{*}}(\pi^{*,\epsilon}) - \beta^{i^{*}}\right),$$
where $i^{*} \in Arg\max_{i}(g^{i}(\pi) - \beta^{i})$.
On rearranging terms in l.h.s:
\begin{equation}
\begin{split}    
\max_{i}\left(g^{i}(\pi^{*,\epsilon}) - \beta^{i}\right) = g^{i^{*}}(\pi^{*, \epsilon}) - \beta^{i^{*}}\leq \frac{\epsilon}{1 - N\epsilon}\sum_{i \neq i^{*}} \beta^{i} - g^{i}(\pi^{*, \epsilon})  \leq \frac{\epsilon}{1 - N\epsilon}\sum_{i}\left(\beta^{i} - \min_{(s,a) \in S\times A }c^{i}(s,a) \right) = \mathcal{O}(\epsilon).
\end{split}
\end{equation}
We get the second inequality by observing that the time average cost for any joint policy is a convex combination of the per-stage costs.
\end{proof}
\subsection{Convergence}
Even though while stating the problem we mentioned that we do not necessarily want convergence, but want to make the joint policy approach a desired set of policies which satisfy the asymptotic bound, in this case we can prove convergence of the algorithm to the policy described in the saddle point above.

\begin{theorem}
$$ \pi_{n} \to \pi^{*,\epsilon}\,\, \text{a.s. for } \epsilon \in (0,1).$$
\end{theorem}
\begin{proof}
Consider the global Q-iterates defined as follows ($\tilde{a}$ is the joint action):
$$ \Tilde{Q}(s,\tilde{a}) = \sum_{i=1}^{N}w(i)Q^{i}(s,\tilde{a}^{i}).$$
The Q-iteration \eqref{eq:Q-learning} for each agent tracks the following ODE ($a^{i}$ is the action taken by the $i$'th agent):
\begin{equation*}
\begin{split}
\dot{Q}^{i}_{t}(s,a^{i}) = \sum_{k=1}^{N}w_{t}(k)c^{k}(s,a^{k}) + \sum_{s'}p(s'|s,a^{i})\min_{b^{i}}Q^{i}_{t}(s',b^{i})  - Q^{i}_{t}(s_{0}, a_{0}^{i}) - Q^{i}_{t}(s,a^{i}).
\end{split}
\end{equation*}
Multiplying the above equation with $w_{t}(i)$ and summing over the agents  $i = 1$ to $N$:
\begin{equation*}
\begin{split}
\dot{\tilde{Q}}_{t}(s,\tilde{a}) = \sum_{k=1}^{N}w_{t}(k)c^{k}(s,a^{k}) + \sum_{s'}p(s'|s, \tilde{a})\min_{\tilde{b}}\tilde{Q}_{t}(s',\tilde{b}) - \tilde{Q}_{t}(s_{0},\tilde{a}_{0}) - \tilde{Q}_{t}(s,\tilde{a}).
\end{split}
\end{equation*}
Treating iterates of a faster (slower) timescale as quasi-equilibriated (quasi-static) is standard in the multiple timescale stochastic approximation theory \citep{borkar2009stochastic-book}. Since $w_{t}$ is modulated on the faster timescale by either the MWU or MH, the above Q-iterations view these as quasi-equilibriated, thus we can drop the $t$-dependence of these weights from the above equations.
Since both the MWU and MH concentrate on the agent $j$ with the highest $z^{j}_{n} - \beta^{j}$, which tracks $g^{j}(\pi_{n}) - \beta^{j}$, the weights converge to $w^{*, \epsilon}$ defined in Theorem \ref{thm : saddle-point}. Thus the global Q-learning optimizes over the joint action space for the per-stage cost $c'(s,\tilde{a})$:
$$ c'(s,\tilde{a}) =\sum_{i=1}^{N}w^{*, \epsilon}(i)c^{i}(s,a^{i}).$$
The global Q-learning iteration converges a.s. to the following policy:
\begin{equation*}
\begin{split}
Arg\min_{\pi \in \Pi}\limsup_{T \to \infty}\frac{1}{T}\sum_{t=1}^{T}\mathbb{E}_{\pi}[c'(X_{t}, \pi(X_{t}))] = Arg\min_{\pi \in \Pi}\sum_{i=1}^{N}w^{*, \epsilon}(i)g^{i}(\pi)  = Arg\min_{\pi \in \Pi}\sum_{i=1}^{N}w^{*, \epsilon}(i)\left(g^{i}(\pi) - \beta^{i}\right) = \pi^{*, \epsilon}.
\end{split}
\end{equation*}
\end{proof}

\section{Stability of the Algorithm}
Until now, we assumed that the iterates of all the stochastic approximation schemes will remain bounded. We prove it in this section.
\subsection{Stability of Gossip}
Since the Gossip iteration \eqref{eq:gossip} averages iterates of the neighbours, for each agent, and the per-stage costs can be assumed to be bounded, the stability is trivial.
\subsection{Stability of Running Costs}
We take a convex combination of the current iterate with the new per-stage cost. Since the per-stage cost can be assumed to be bounded, we can prove by induction that the iterates in \eqref{eq:running-cost} will also remain bounded.
\subsection{Stability of Q-learning}
We shall use the stability test that was stated in the preliminaries.
The limiting ODE for the relative Q-value iteration \eqref{eq:Q-learning} of agent $i$ is:
$$ \Dot{Q}^{i}(t) = T(Q^{i}_{w}(t)) - Q^{i}(t),$$
$$T(Q^{i}_{w}(t))(s,a) = \sum_{j}w(j) c^{j}(s,a) - Q^{i}(s_{0},a_{0}) + \sum_{s'}p(s'|s, \bar{a}')\min_{b}Q^{i}(s',b),$$
where $\bar{a}$ is the joint action of the agents.
Due to asynchronicity, the ODE becomes:
$$ \Dot{Q}^{i}(t) = \Lambda^{i}(t)[T(Q^{i}_{w}(t)) - Q^{i}(t)] = \hat{T}^{i}_{w, \Lambda^{i}(t)}[Q^{i}(t)],$$
$\Lambda^{i}(t)$ is a diagonal matrix with elements in $[0,1]$ representing the relative frequency of updates.\\
Consider the following ODE:
\begin{equation}\label{eq : stability test ODE}
\Dot{Q}^{i}(t) = \hat{T}^{i, \infty}_{w, \Lambda^{i}(t)}[Q^{i}(t)],
\end{equation}
where we  define the scaled operator as follows:
$$ \hat{T}^{i, \infty}_{w, \Lambda^{i}(t)}[q] = \lim_{c \to \infty}\frac{\hat{T}^{i}_{w, \Lambda^{i}(t)}[c q]}{c}.$$
Since the origin is the unique globally asymptotically stable equilibrium of \eqref{eq : stability test ODE} uniformly in the external input $w$ (which remains bounded due to boundedness of the Gossip iteration \eqref{eq:gossip}) and uniformly in $\Lambda^{i}(t)$, with the diagonal elements of $\Lambda^{i}(t)$ being bounded away from $0$ (due to Assumption 4), the Q-iterates for all the agents remain bounded. The rest of the arguments are the same as \citep{BHATNAGAR2011472, asynchronous-ODE}.

\section{Numerical Simulations}
In this section, we demonstrate the performance of both variants (MWU and MH) of our algorithm in different environments. We first show the performance in an environment where the per-stage costs of each agent depends only on their individual actions (the assumption under which we guarantee performance), we call this the "simple" case. We then show that our algorithm empirically works even in the more general setting where the per-stage costs of each agent depends on the joint actions of all the agents. We call this the "general" case.
\label{ch:name} 

\subsection{7-Agent 2-state-2-action MDP (Simple case)}
\begin{figure}[!ht]
    \centering
    \includegraphics[width= 10cm]{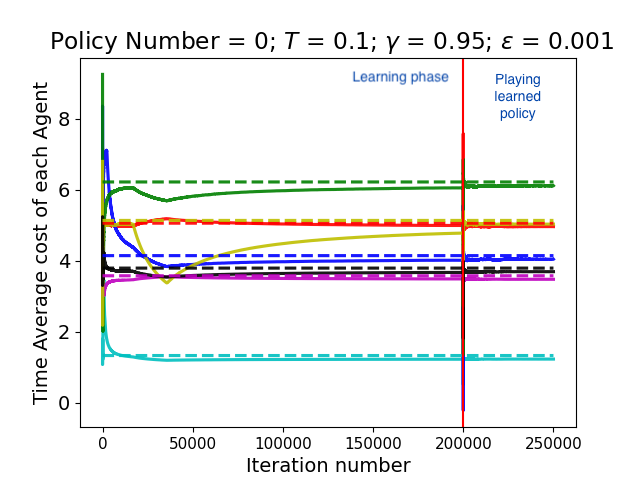}
    \caption{7-Agent 2-state-2-action MDP (simple case) with Multiplicative Weights Update - Policy Number $0$}
    \label{fig-1}
\end{figure}

\begin{figure}[!ht]
    \centering
    \includegraphics[width= 10cm]{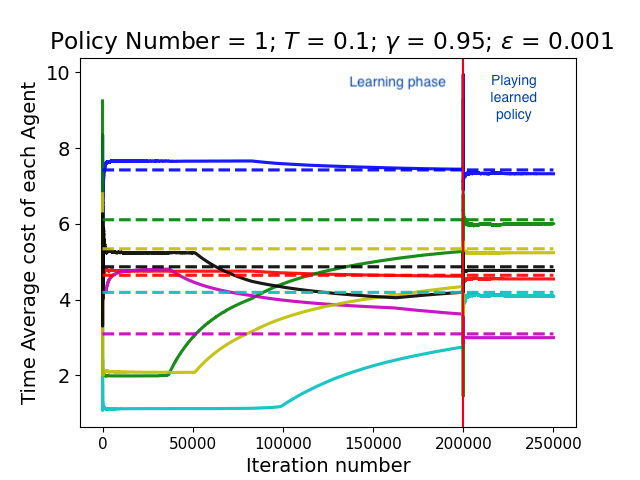}
    \caption{7-Agent 2-state-2-action MDP (simple case) with Multiplicative Weights Update - Policy Number $1$}
    \label{fig-2}
\end{figure}

\begin{figure}[!ht]
    \centering
    \includegraphics[width= 10cm]{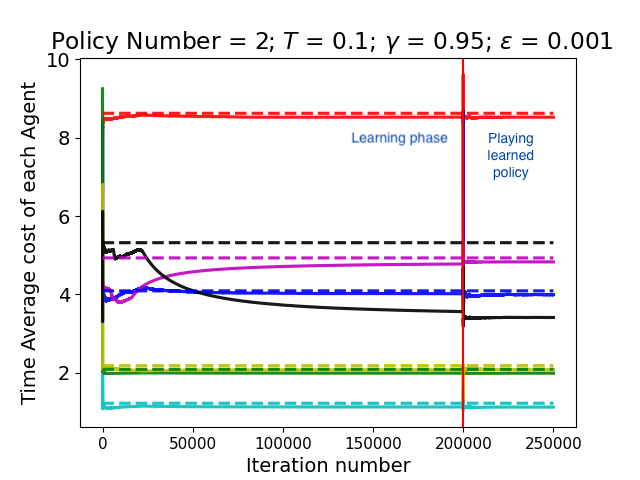}
    \caption{7-Agent 2-state-2-action MDP (simple case) with Multiplicative Weights Update - Policy Number $2$}
    \label{fig-3}
\end{figure}

\begin{figure}[!ht]
    \centering
    \includegraphics[width= 10cm]{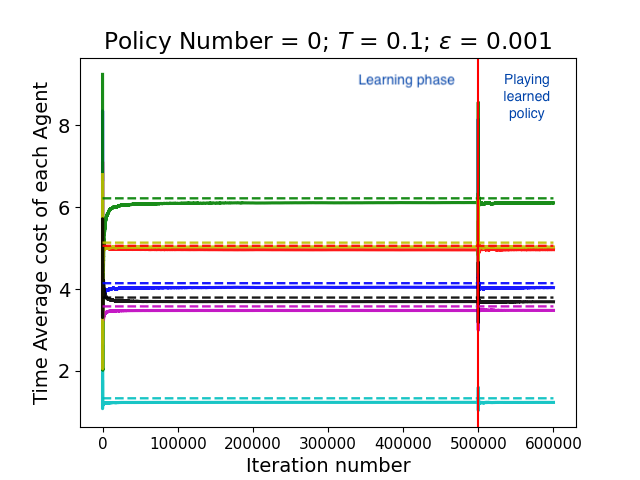}
    \caption{7-Agent 2-state-2-action MDP (simple case) with Metropolis-Hastings scheme - Policy Number $0$}
    \label{fig-4}
\end{figure}

\begin{figure}[!ht]
    \centering
    \includegraphics[width= 10cm]{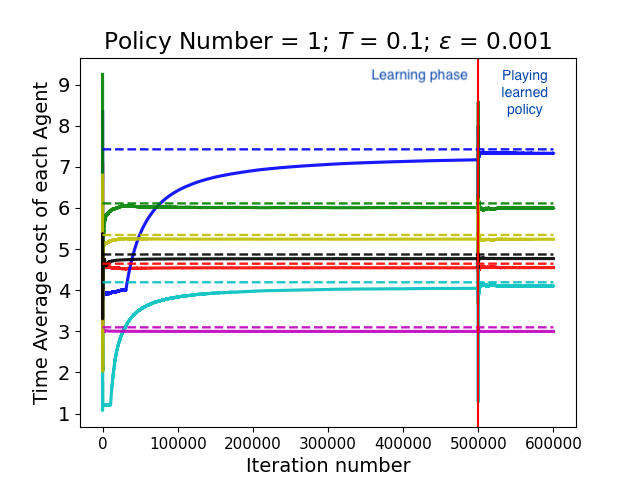}
    \caption{7-Agent 2-state-2-action MDP (simple case) with Metropolis-Hastings scheme - Policy Number $1$ }
    \label{fig-5}
\end{figure}

\begin{figure}[!ht]
    \centering
    \includegraphics[width= 10cm]{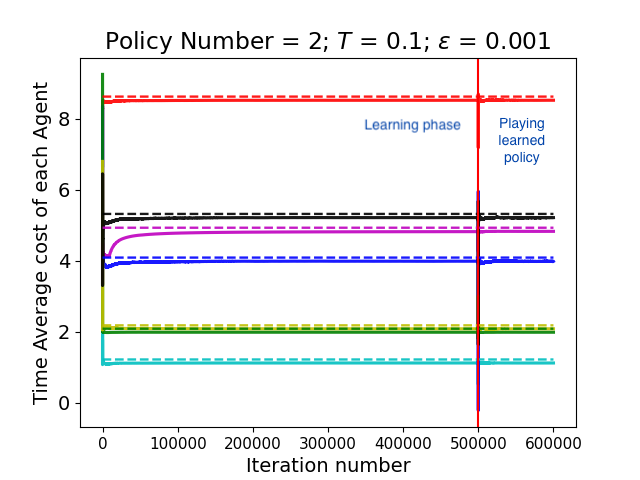}
    \caption{7-Agent 2-state-2-action MDP (simple case) with Metropolis-Hastings scheme - Policy Number $2$}
    \label{fig-6}
\end{figure}

We have an MDP which has 2 states $(s_{0}, s_{1})$  and 2 actions $(0,1)$ and we use it as a multi-agent MDP by taking XOR of the actions of each agent. The transition kernel and cost functions are randomly generated, with the per-stage costs being drawn uniformly at random from $[0,10]$. We set the value of $\beta$ as $0.1$ above the time average for each agent corresponding to some of the available policies. We stop the learning after some iterations and simply run the learned policy (greedy w.r.t. learned Q-values) for the rest of the iterations to demonstrate that the required bounds are indeed satisfied.
The plots are given in Figures \ref{fig-1}-\ref{fig-6}.
\subsection{7-Agent 10-state-2-action MDP (Simple case)}
\begin{figure}[!ht]
    \centering
    \includegraphics[width= 10cm]{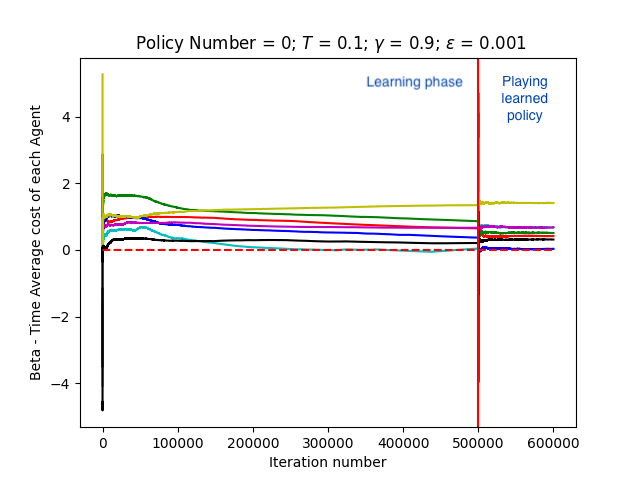}
    \caption{7-Agent 10-state-2-action MDP (simple case) with Multiplicative Weights Update - Policy Number $0$}
    \label{fig-7}
\end{figure}

\begin{figure}[!ht]
    \centering
    \includegraphics[width= 10cm]{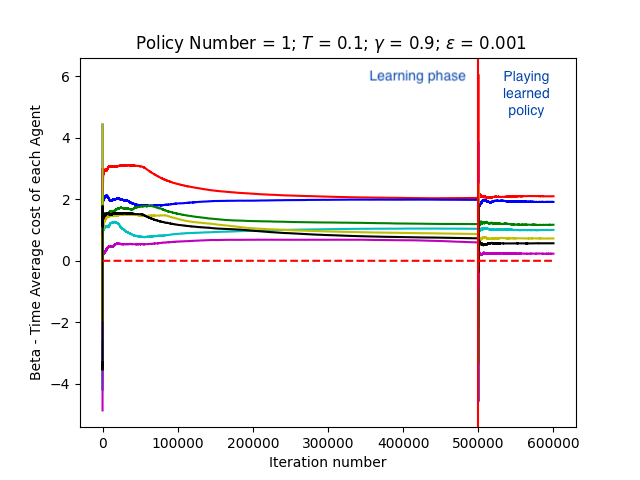}
    \caption{7-Agent 10-state-2-action MDP (simple case) with Multiplicative Weights Update - Policy Number $1$}
    \label{fig-8}
\end{figure}

\begin{figure}[!ht]
    \centering
    \includegraphics[width= 10cm]{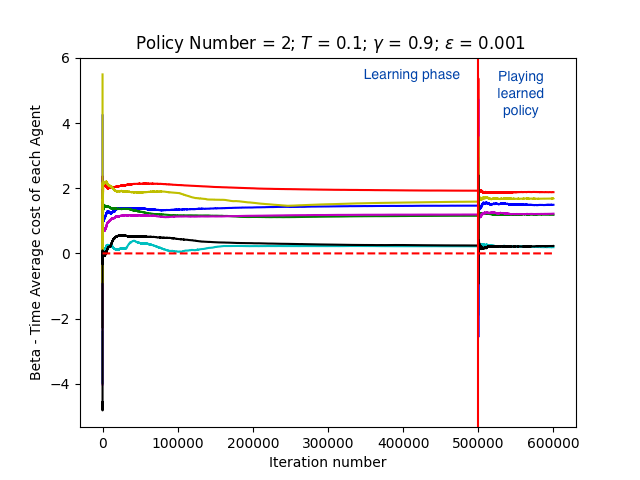}
    \caption{7-Agent 10-state-2-action MDP (simple case) with Multiplicative Weights Update - Policy Number $2$}
    \label{fig-9}
\end{figure}

\begin{figure}[!ht]
    \centering
    \includegraphics[width= 10cm]{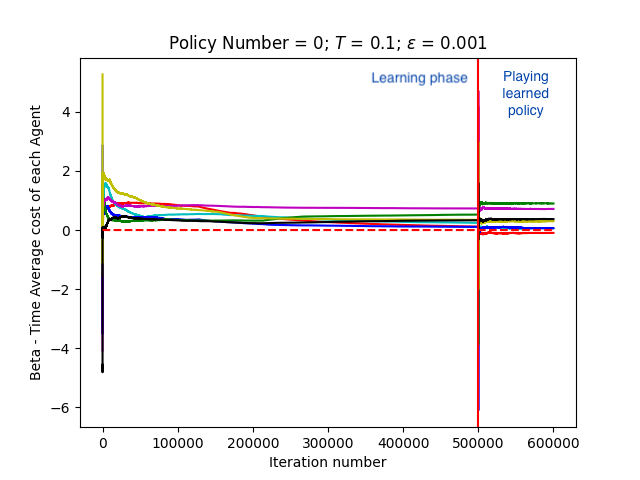}
    \caption{7-Agent 10-state-2-action MDP (simple case) with Metropolis-Hastings scheme - Policy Number $0$}
    \label{fig-10}
\end{figure}

\begin{figure}[!ht]
    \centering
    \includegraphics[width= 10cm]{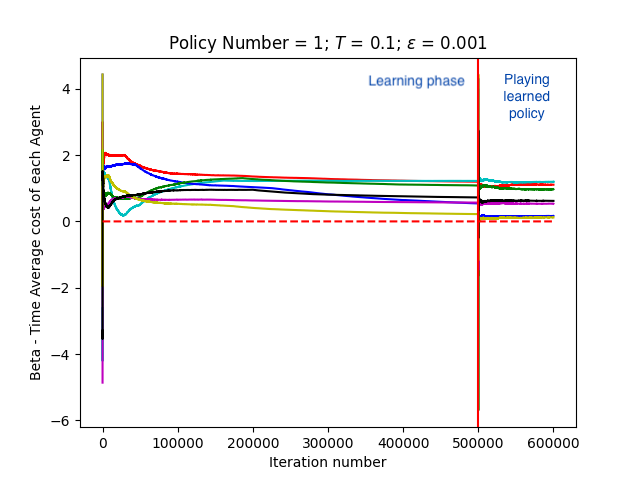}
    \caption{7-Agent 10-state-2-action MDP (simple case) with Metropolis-Hastings scheme - Policy Number $1$}
    \label{fig-11}
\end{figure}

\begin{figure}[!ht]
    \centering
    \includegraphics[width= 10cm]{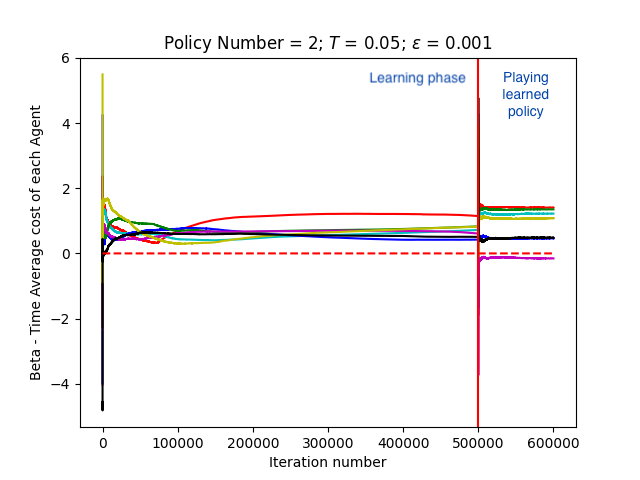}
    \caption{7-Agent 10-state-2-action MDP (simple case) with Metropolis-Hastings scheme - Policy Number $2$}
    \label{fig-12}
\end{figure}

We now repeat the above experiments with an MDP with 10 states and 2 actions . However, in these plots we make the y-axis $\beta^{i} - z^{i}$, where $\beta^{i}$ is the desired asymptotic bound on the the time average cost of agent $i$ and $z^{i}$ is it's time average cost. This is to increase the readability of these graphs, since the curves may be very close to each other. We expect that the curves, for each agent, should lie above the x-axis (marked by the horizontal dotted red line). The plots are given in Figures \ref{fig-7} - \ref{fig-12}.

\subsection{7-Agent 2-state-2-action MDP (General case)}
\begin{figure}[!ht]
    \centering
    \includegraphics[width= 10cm]{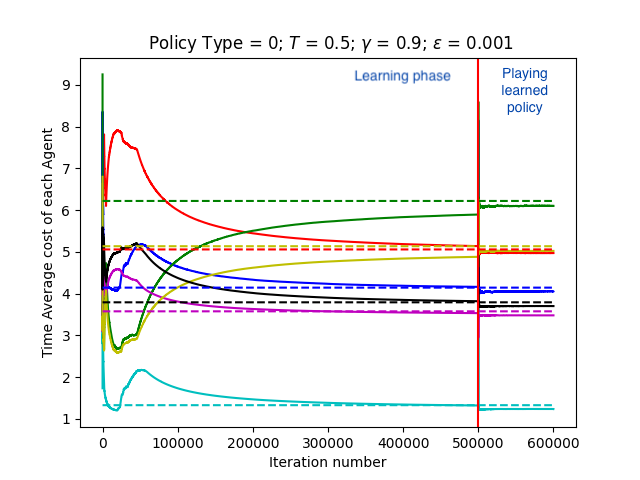}
    \caption{7-Agent 2-state-2-action MDP (general case) with Multiplicative Weights Update - Policy Number $0$}
    \label{fig-13}
\end{figure}

\begin{figure}[!ht]
    \centering
    \includegraphics[width= 10cm]{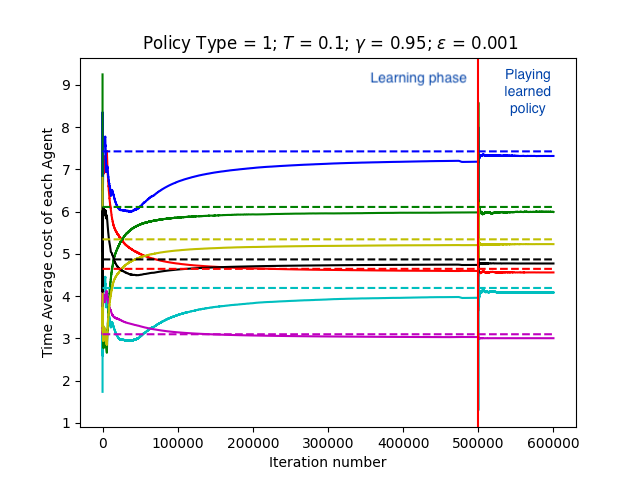}
    \caption{7-Agent 2-state-2-action MDP (general case) with Multiplicative Weights Update - Policy Number $1$}
    \label{fig-14}
\end{figure}

\begin{figure}[!ht]
    \centering
    \includegraphics[width= 10cm]{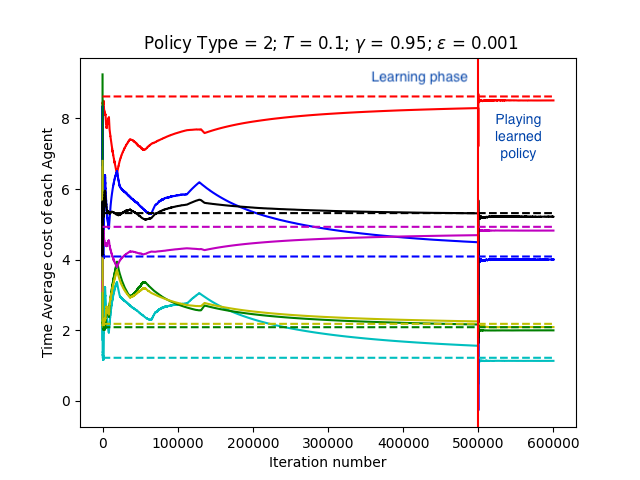}
    \caption{7-Agent 2-state-2-action MDP (general case) with Multiplicative Weights Update - Policy Number $2$}
    \label{fig-15}
\end{figure}

\begin{figure}[!ht]
    \centering
    \includegraphics[width= 10cm]{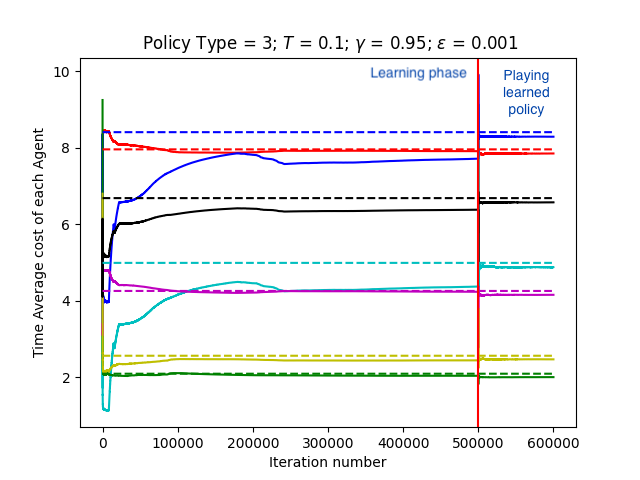}
    \caption{7-Agent 2-state-2-action MDP (general case) with Multiplicative Weights Update - Policy Number $3$}
    \label{fig-16}
\end{figure}

\begin{figure}[!ht]
    \centering
    \includegraphics[width= 10cm]{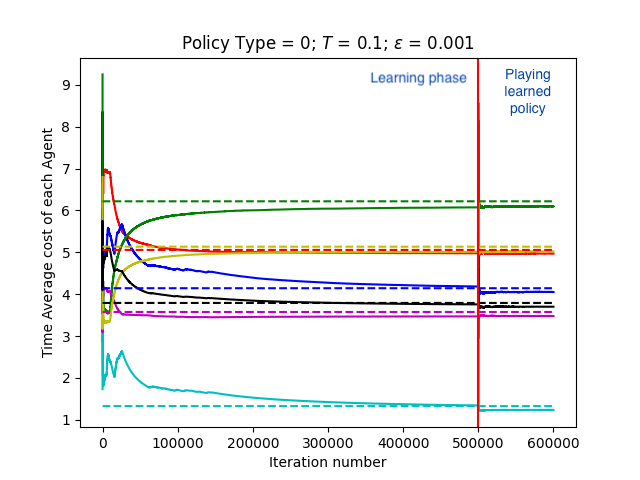}
    \caption{7-Agent 2-state-2-action MDP (general case) with Metropolis-Hastings scheme - Policy Number $0$}
    \label{fig-17}
\end{figure}

\begin{figure}[!ht]
    \centering
    \includegraphics[width= 10cm]{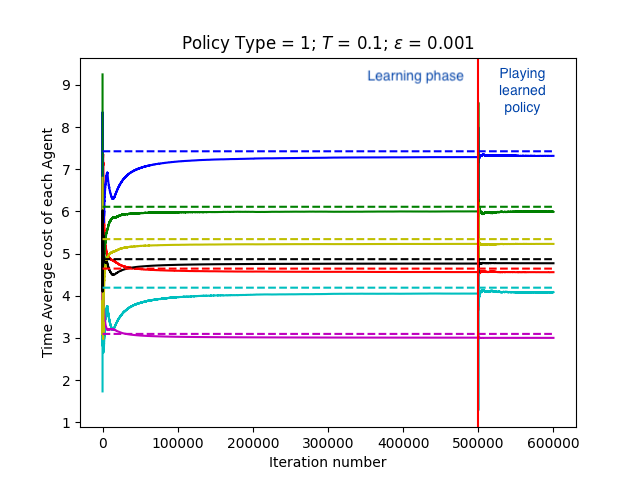}
    \caption{7-Agent 2-state-2-action MDP (general case) with Metropolis-Hastings scheme - Policy Number $1$}
    \label{fig-18}
\end{figure}

\begin{figure}[!ht]
    \centering
    \includegraphics[width= 10cm]{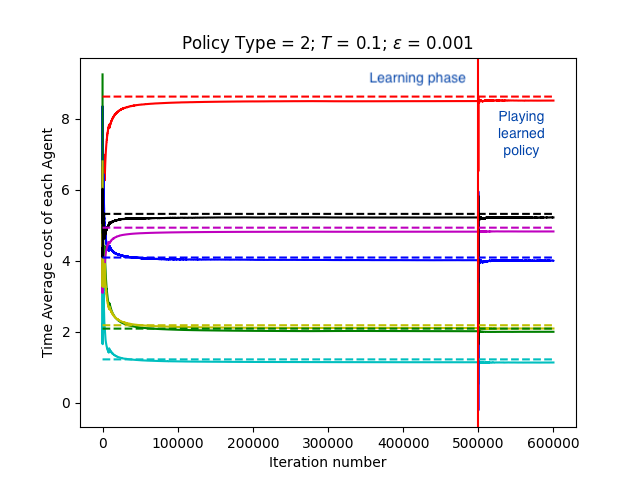}
    \caption{7-Agent 2-state-2-action MDP (general case) with Metropolis-Hastings scheme - Policy Number $2$}
    \label{fig-19}
\end{figure}

\begin{figure}[!ht]
    \centering
    \includegraphics[width= 10cm]{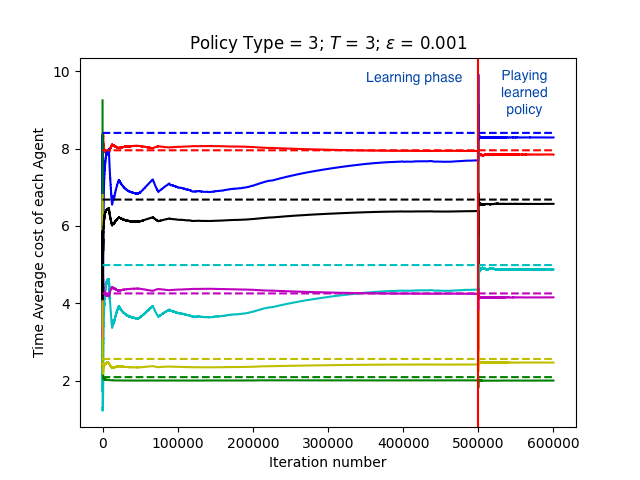}
    \caption{7-Agent 2-state-2-action MDP (general case) with Metropolis-Hastings scheme - Policy Number $3$}
    \label{fig-20}
\end{figure}

The setting is the same as previously described, but here the per-stage costs of each agent depend on the joint-actions of all the agents. The plots are in Figures \ref{fig-13}-\ref{fig-20}.

\subsection{4-Agent Queueing system (General case)}

\begin{figure}[!ht]
    \centering
    \includegraphics[width= 10cm]{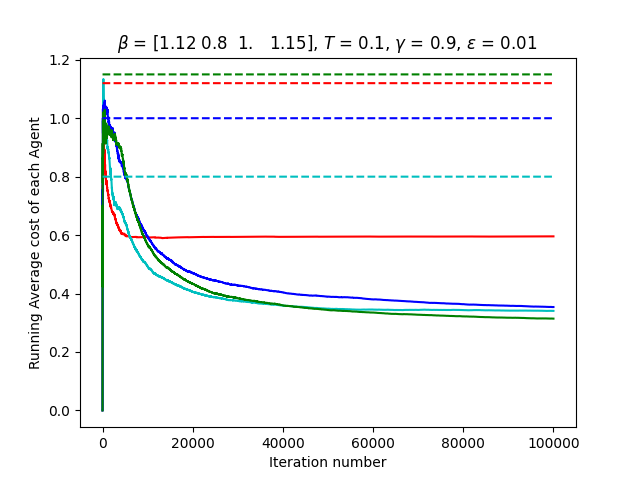}
    \caption{4-Agent queueing system with Multiplicative Weights Update - 0}
    \label{fig-21}
\end{figure}

\begin{figure}[!ht]
    \centering
    \includegraphics[width= 10cm]{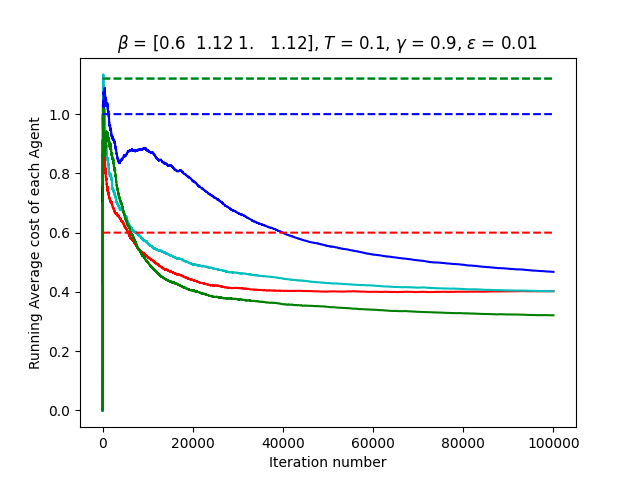}
    \caption{4-Agent queueing system with Multiplicative Weights Update - 1}
    \label{fig-22}
\end{figure}

\begin{figure}[!ht]
    \centering
    \includegraphics[width= 10cm]{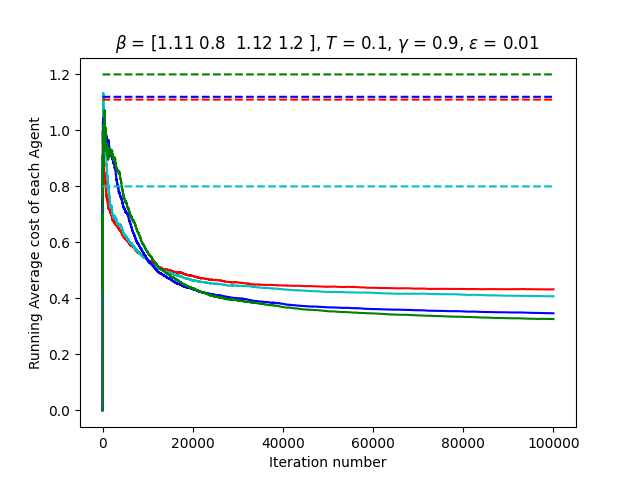}
    \caption{4-Agent queueing system with Multiplicative Weights Update - 2}
    \label{fig-23}
\end{figure}

\begin{figure}[!ht]
    \centering
    \includegraphics[width= 10cm]{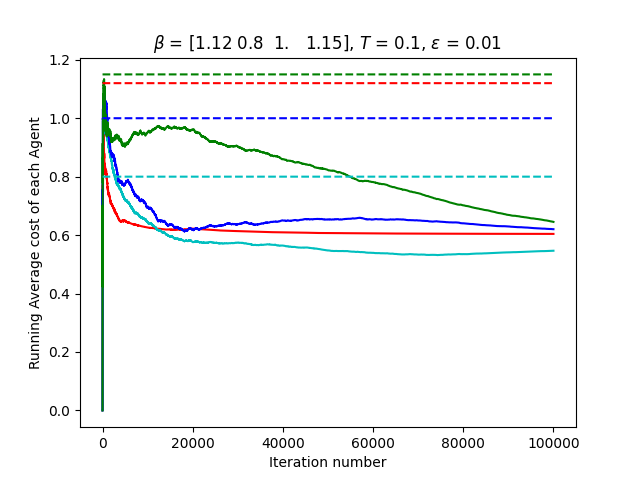}
    \caption{4-Agent queueing system with Metropolis-Hastings scheme - 0}
    \label{fig-24}
\end{figure}

\begin{figure}[!ht]
    \centering
    \includegraphics[width= 10cm]{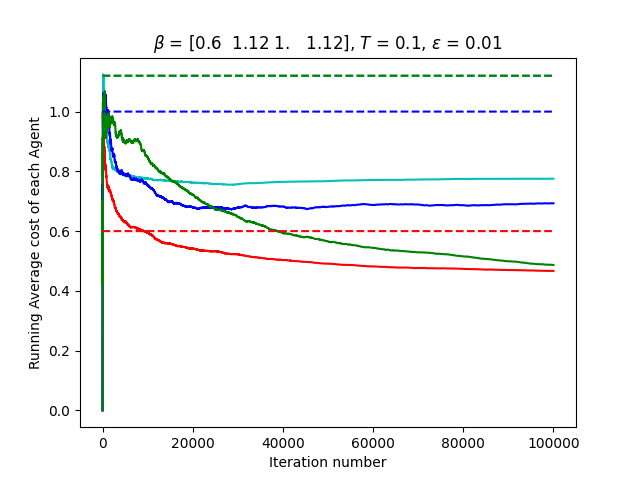}
    \caption{4-Agent queueing system with Metropolis-Hastings scheme - 1}
    \label{fig-25}
\end{figure}

\begin{figure}[!ht]
    \centering
    \includegraphics[width= 10cm]{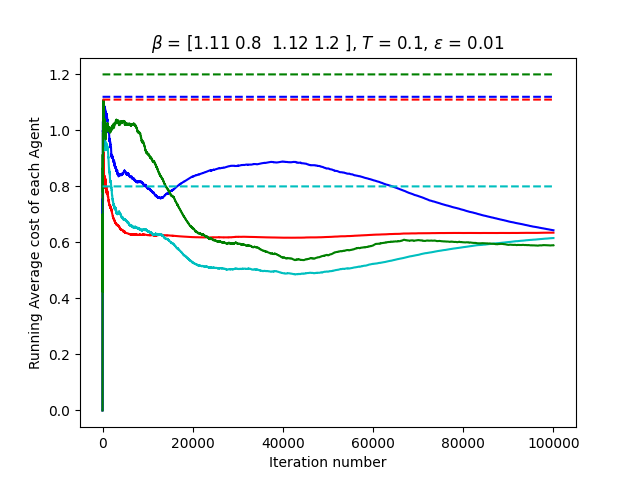}
    \caption{4-Agent queueing system with Metropolis-Hastings scheme - 2}
    \label{fig-26}
\end{figure}

We simulate a 4-agent queue for sending data packets over a common channel, with a buffer size of 2 for each agent's queue. The state is given by the tuple $(X^{0}, X^{1}, X^{2}, X^{3})$ where $X^{i}$ is the number of packets in agent $i$'s queue. The arrival of data packets for each queue follows a Bernoulli process.
From each state, agents can choose whether to transmit exactly one packet or not transmit at all. The state dynamics is given by:
$$X_{n+1}^{i} = X^{i}_{n} - \xi_{n}^{i}\mathbb{I}(X^{i}_{n}\geq 1) + A_{n}^{i},$$
where $\xi_{n}^{i}$ is the action of agent $i$, and $A_{n}^{i}$ for all $i$ are Bernoulli processes, with $A_{n}^{i} \sim \text{Ber}(p_{i})$.
The cost of each agent is:
$$ c^{i}(\Bar{X}_{n}, \Bar{\xi}_{n}) = \mu_{n}\xi_{n}^{i}\prod_{j: j\neq i}(1-\xi_{n}^{j}) + K \xi_{n}^{i}\sum_{j\neq i}\xi_{n}^{j} + l^{i}X_{n}^{i},$$
where $\Bar{X}_{n} = (X^{0}_{n}, X^{1}_{n}, X^{2}_{n}, X^{3}_{n})$ and $\Bar{\xi}_{n} = (\xi^{0}_{n}, \xi^{1}_{n}, \xi^{2}_{n}, \xi^{3}_{n})$, $\mu_{n}$ is the cost of transmitting over the channel (which may depend on some time varying SNR), which is $\mu_{h}$(high) w.p. $p_{h}$ and $\mu_{l}$(low) otherwise, $K$ is the cost of collision and $l^{i}$ is the per-stage cost for maintaining packets in agent $i$'s queue (which should be proportional to the delay experienced). Note that $K > \mu_{h}p_{h} + \mu_{l}(1-p_{h})$ i.e. on average, a collision costs more than transmission.\\
In the numerical experiments, we use : $p_{0} = 0.35, p_{1} = 0.3, p_{2} = 0.25, p_{3} = 0.2.$ for the Bernoulli arrival processes of each queue. The collision cost $K$ is $1$, the cost of transmitting is $0.2$ (high SNR) and $0.8$ (low SNR) with equal probability. The per-stage costs of maintaining data packets in the queue are $l_{0} = 0.3, l_{1} =0.4 , l_{2} = 0.5, l_{3} = 0.6$. The plots are given in Figures \ref{fig-21} - \ref{fig-26}.

\subsection{2-Agent Grid world environment (General case)}

\begin{figure}[!ht]
    \centering
    \includegraphics[width= 10cm]{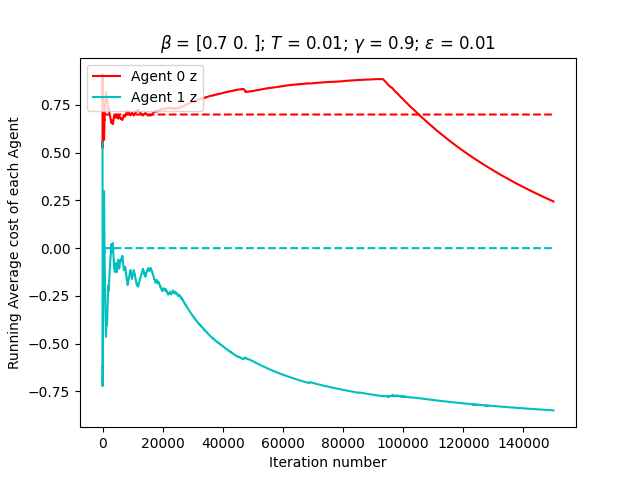}
    \caption{2-agent grid world with the Multiplicative Weights Update - 0}
    \label{fig-27}
\end{figure}

\begin{figure}[!ht]
    \centering
    \includegraphics[width= 10cm]{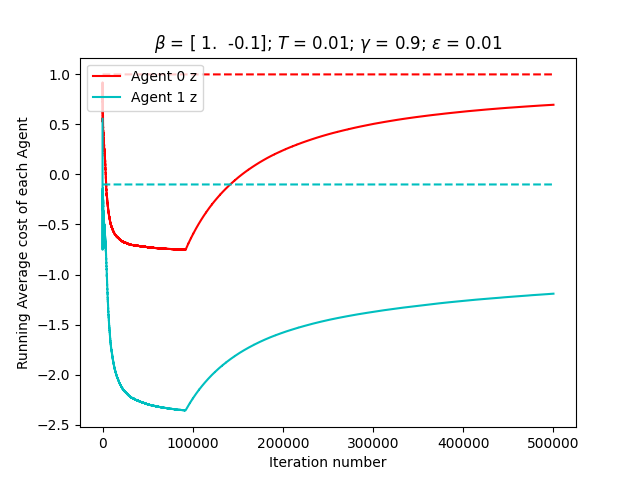}
    \caption{2-agent grid world with the Multiplicative Weights Update - 1}
    \label{fig-28}
\end{figure}

\begin{figure}[!ht]
    \centering
    \includegraphics[width= 10cm]{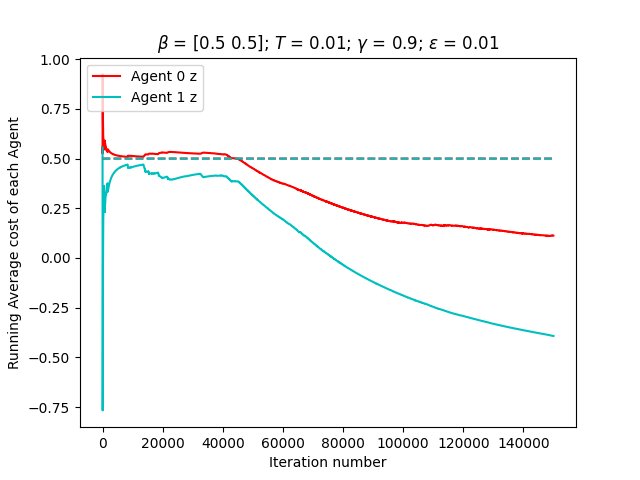}
    \caption{2-agent grid world with the Multiplicative Weights Update - 2}
    \label{fig-29}
\end{figure}

\begin{figure}[!ht]
    \centering
    \includegraphics[width= 10cm]{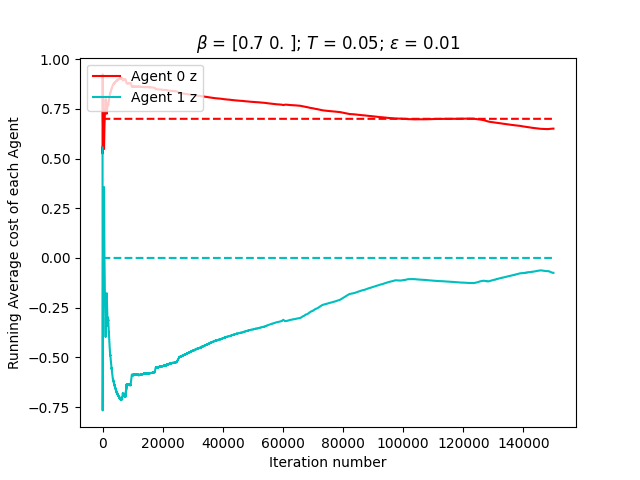}
    \caption{2-agent grid world with the Metropolis-Hastings scheme - 0}
    \label{fig-30}
\end{figure}

\begin{figure}[!ht]
    \centering
    \includegraphics[width= 10cm]{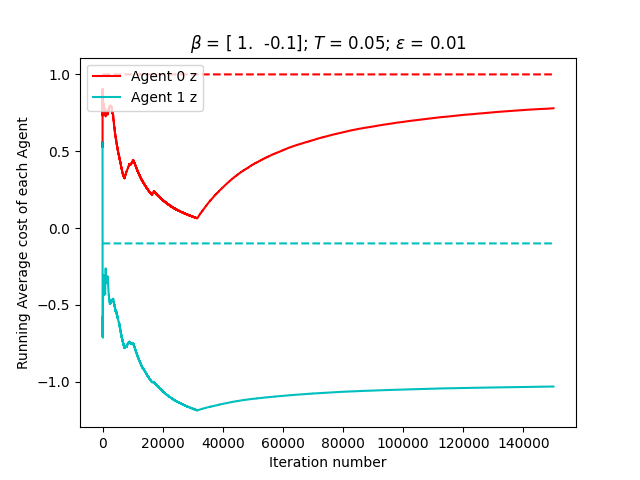}
    \caption{2-agent grid world with the Metropolis-Hastings scheme - 1}
    \label{fig-31}
\end{figure}

\begin{figure}[!ht]
    \centering
    \includegraphics[width= 10cm]{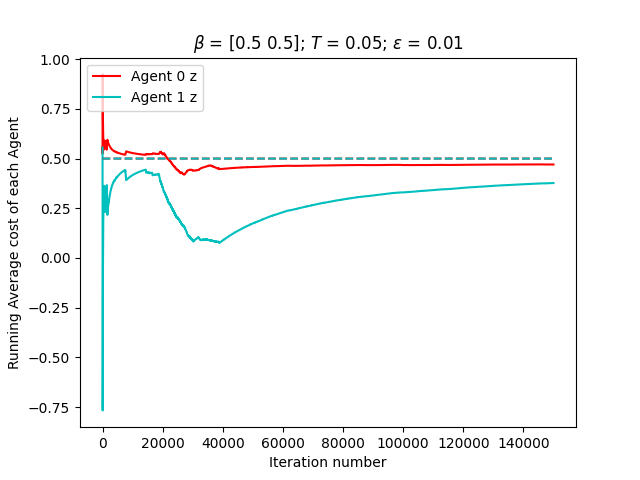}
    \caption{2-agent grid world with the Metropolis-Hastings scheme - 2}
    \label{fig-32}
\end{figure}

We modify the environment in \citet{jiang2021multi} for this experiment.
We have a 2-agent grid $(6\times 6)$ environment, where both the agents start at the bottom left and need to navigate to the top right of the grid. From each state, given by the positions of the 2 agents, they can move either up, down, left or right. Reaching the top right of the grid together gives both agents a reward of 10 and resets the environment. When both the agents are not in the top right, a per-stage cost of 0.5 is incurred by both the agents. If the distance between the two agents is less than or equal 1, one of them incurs a cost of 1 and the other gets a reward of 1, instead of both getting 0.5. The plots are given in Figures \ref{fig-27} - \ref{fig-32}.

\section{Conclusions}
We considered Multi-agent MDPs in which the transitions are controlled by the actions of all the agents and each agent has its own cost function controlled by its own actions. Inspired by Blackwell approachability, the goal was to make the time averaged costs of each of the agents satisfy certain asymptotic bounds. We suggested two variants of an algorithm, one which uses Multiplicative Weights Update and the other which uses the Metropolis Hastings algorithm. We proved that both the variants satisfy the desired asymptotic bounds approximately. We also showed that the proposed algorithm has good empirical performance even in the more general case, when the costs depend on the joint-actions of all agents. Future works could include investigating convergence properties for this algorithm in the more general setting and also searching for meta-algorithms that could replace the Multiplicative weights update and Metropolis-Hastings scheme.  Since stochastic iterations such as the ones used in our algorithm can be slow, an analysis of convergence rate to the desired set of policies as well as procedures to speed this up would be desirable for practical purposes. It would be interesting to see a finite time analysis of such algorithms that ensure convergence to some set of policies.
\section{Figures}
This section includes the plots for the numerical simulations previously described.


\clearpage

\bibliographystyle{unsrtnat}
\bibliography{references}

\end{document}